\documentclass[conference]{IEEEtran}
\usepackage{amssymb}
\usepackage{color}
\newtheorem{theorem}{Theorem}
\newtheorem{lemma}{Lemma}
\newtheorem{corollary}{Corollary}

\usepackage{cite}
\ifCLASSINFOpdf
  \usepackage[pdftex]{graphicx}
  \graphicspath{{../pdf/}{../jpeg/}}
  \DeclareGraphicsExtensions{.pdf,.jpeg,.png}
\else
  \usepackage[dvips]{graphicx}
  \graphicspath{{../eps/}}
  \DeclareGraphicsExtensions{.eps}
\fi
\usepackage[cmex10]{amsmath}
\usepackage{array}

\usepackage[tight,footnotesize]{subfigure}
\usepackage{microtype}
\begin{document}
\title{The Capacity of Three-Receiver AWGN Broadcast Channels with Receiver Message Side Information}


\author{\IEEEauthorblockN{Behzad Asadi, Lawrence Ong, and Sarah J.\ Johnson}
\IEEEauthorblockA{School of Electrical Engineering and Computer Science, The University of Newcastle, Newcastle, Australia} Email:{ behzad.asadi@uon.edu.au, lawrence.ong@cantab.net, sarah.johnson@newcastle.edu.au}}

\maketitle

\begin{abstract}
This paper investigates the capacity region of three-receiver AWGN broadcast channels where the receivers (i) have private-message requests and (ii) know the messages requested by some other receivers as side information. We classify these channels based on their side information into eight groups, and construct different transmission schemes for the groups.
For six groups, we characterize the capacity region, and show that it improves both the best known inner and outer bounds. For the remaining two groups, we improve the best known inner bound 
by using side information during channel decoding at the receivers.
\end{abstract}


%
\IEEEpeerreviewmaketitle

\section{Introduction}
We study the capacity region of three-receiver additive white Gaussian noise broadcast channels (AWGN BCs) where the receivers have private-message requests and know some of the transmitted messages, aimed for other receivers, a priori.
\subsection{Background}
Broadcast channels \cite{BC} are considered as one of the main components of multi-sender multi-receiver wireless networks. The capacity region of broadcast channels is not known in general, except for a few special classes, e.g., degraded broadcast channels, which include AWGN BCs~\cite{AWGNBCConverse}. 

A variant of broadcast channels is where the receivers have some information about the source messages a priori (referred to as receiver message side information). 
This models several practical applications, e.g., sensor networks where the receivers know noisy versions of the source messages~\cite{SWoverBC}.
For some applications, e.g., multimedia broadcasting with packet loss or the downlink phase of multi-way relay channels, the receivers know some noise-free parts of the source messages. 

The capacity region of broadcast channels with receiver message side information where each receiver must decode all the source messages (or equivalently, all the messages not known a priori) has been established by Tuncel~\cite{SWoverBC} and Oechtering et al.~\cite{BCwithSI2UsersOechtering}.

However, the case where the receivers need not decode all the messages remains unsolved to date. Wu characterized the capacity region of \textit{two-receiver} AWGN BCs with general message request and receiver message side information~\cite{BCwithSI2UsersGeneral}. 
Extending the results to three or more receivers is ``highly nontrivial''~\cite{BCwithSI2UsersGeneral}. 
Oechtering et al.\ characterized the capacity region of some classes of \textit{three-receiver} less-noisy and more-capable broadcast channels, where (i) only two receivers possess side information and (ii) the request of the third receiver is only restricted to a common message~\cite{BCwithSI3UsersCommonMessage}.


\subsection{Existing Results and Contributions}
In this paper, we consider \textit{private-message} broadcasting over three-receiver AWGN BCs where the receivers know the messages requested by some other receivers as side information.  
The best known inner and outer bounds are within a constant gap of the capacity region~\cite{BCwithSI3UsersPrivateMessage}; the inner bound (achievability) uses a separate index and channel coding scheme, developed based on the deterministic approach\cite{Deterministic}. 

One of the difficulties in deriving the capacity region is to find a unified scheme for all side information configuations.
To make the problem more tractable, we first classify the channels into eight groups based on their side information, and construct different transmission schemes for different groups. 

\begin{figure}[t]
\centering
\includegraphics[width=0.38\textwidth]{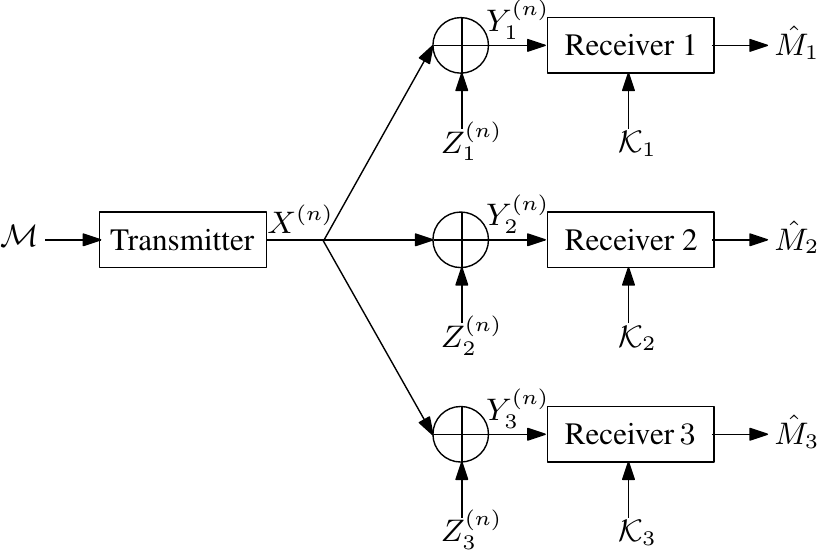}
\vspace{-8pt}
\caption{The AWGN broadcast channel with receiver message side information, where $\mathcal{M}=\{M_1,M_2,M_3\}$ is the set of independent messages, each demanded by one receiver, and  $\mathcal{K}_i \subseteq \mathcal{M}\setminus\{M_i\}$ is the set of messages known to receiver $i$ a priori.} 
\vspace{-12pt}
\label{AWGNBCModelFig}
\end{figure}

For six groups, we establish the capacity region. 
Our classification proves to be useful in grouping the channels with the same capacity-achieving transmission scheme.
This result also shows the looseness of the best known inner and outer bounds~\cite{BCwithSI3UsersPrivateMessage}.
For the remaining two groups, we improve the capacity inner bound by using side information during channel decoding at the receivers.




\section{AWGN BC with Side Infromation}\label{AWGN BC with SI}
In the channel model under consideration, as depicted in Fig. \ref{AWGNBCModelFig}, the signals received by receiver $i$, $Y_{i}^{(n)}=\left(Y_{i1},Y_{i2},\ldots,Y_{in}\right)\;i=1,2,3$, is the sum of the transmitted codeword, $X^{(n)}$, and an i.i.d. noise sequence, $Z_i^{(n)} \;i=1,2,3$, with normal distribution, $Z_i\sim \mathcal{N}\left(0, N_i\right)$. This channel is stochastically degraded, and without loss of generality, we can assume that receiver $1$ is the strongest and receiver $3$ is the weakest in the sense that $N_1\leq N_{2} \leq N_{3}$.

The transmitted codeword has a power constraint of $\sum_{l=1}^{n}E\left(X_l^2\right)\leq nP$ and is a function of source messages $\mathcal{M}=\{M_1,M_2,M_3\}$. The messages $\{M_i\}_{i=1}^3$ are independent, and $M_i$ is intended for receiver $i$ at rate $R_i$ bits per channel use i.e., $m_i\in \{1,2,\ldots,2^{nR_i}\}$.  To model the side information of each receiver, one set is defined corresponding to each receiver; the \textit{knows} set, $\mathcal{K}_i$, is the set of messages known to receiver $i$.

\begin{figure}[t]
\centering
\includegraphics[width=0.15\textwidth]{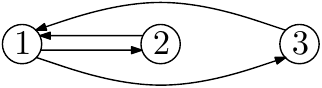}
\vspace{-7pt}
\caption{Sample side information graph where receiver 1 knows $M_2$ and $M_3$, receiver 2 knows $M_1$, and receiver 3 knows $M_1$.} 
\vspace{-5pt}
\label{SampleSIGraph}
\end{figure}

The side information configuration of each channel is modeled by a side information graph, $\mathcal{G}=\left(\mathcal{V}_\mathcal{G},\mathcal{A}_\mathcal{G}\right)$, where $\mathcal{V}_\mathcal{G} = \{1,2,3\}$ is the set of \textit{vertices} and $\mathcal{A}_\mathcal{G}$ is the set of \textit{arcs}. As we have only private messages, vertex $i$ represents both $M_i$ and receiver $i$ requesting it. An arc from vertex $i$ to vertex $j$, denoted by $(i,j)$, exists if and only if receiver $i$ knows $M_j$. The set of out-neighbors of vertex $i$ is then $\mathcal{O}_i\triangleq\{j\mid (i,j)\in\mathcal{A}_\mathcal{G}\}=\{j\mid M_j\in\mathcal{K}_i\}$. A sample side information graph is shown in Fig. \ref{SampleSIGraph}.

\begin{figure}[t]
\centering
\includegraphics[width=0.33\textwidth]{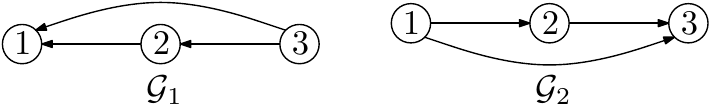}
\vspace{-8pt}
\caption{Defined graphs in order to classify the problem.} 
\vspace{-10pt}
\label{G1G2}
\end{figure}

\begin{figure}[t]
\centering
\includegraphics[width=0.36\textwidth]{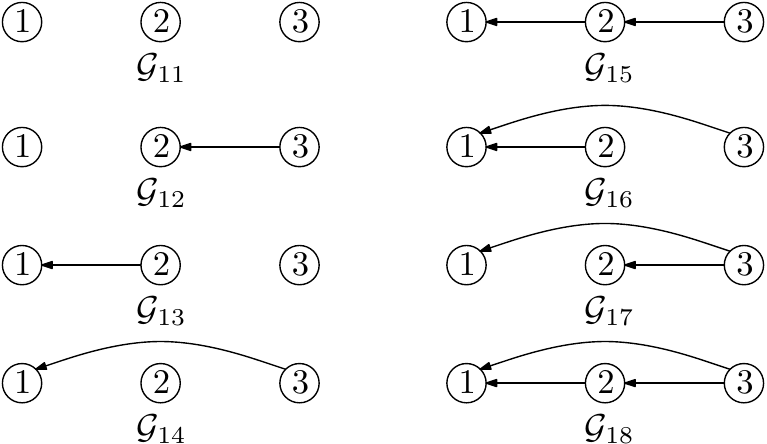}
\vspace{-7pt}
\caption{Group leaders, capturing if each receiver knows the message(s) requested by stronger receiver(s).}
\vspace{-11pt} 
\label{GroupLeaders}
\end{figure}
\begin{figure}[b]
\centering
\includegraphics[width=0.36\textwidth]{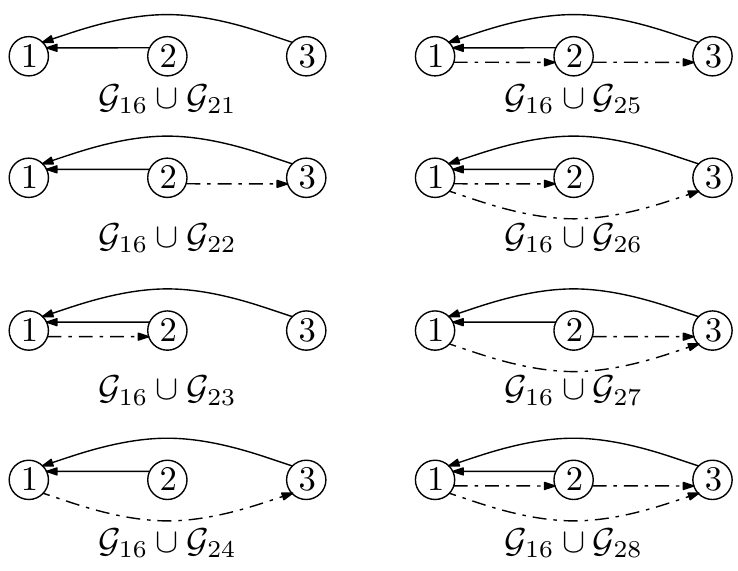}
\vspace{-7pt}
\caption{The elements of group 6, where the arcs of $\mathcal{G}_{16}$ (the group leader) are drawn with solid lines, and those of $\mathcal{G}_{2j}$ dotted lines. As it can be seen, the group leader is actually the first member of each group.} 
\label{ElementsofGroups6}
\end{figure}

\section{Problem Classification}
We classify the channels of interest 
into eight groups based on their side information graphs. To this end, we define two graphs, $\mathcal{G}_1=\left(\mathcal{V}_{\mathcal{G}_1}, \mathcal{A}_{\mathcal{G}_1}\right)$ and $\mathcal{G}_2 = \left(\mathcal{V}_{\mathcal{G}_2}, \mathcal{A}_{\mathcal{G}_2}\right)$ where $\mathcal{V}_{\mathcal{G}_1}=\mathcal{V}_{\mathcal{G}_2}=\mathcal{V}_{\mathcal{G}}$, as shown in Fig.~\ref{G1G2}. 
A side information graph is the union of an arc subgraph\footnote{ $\mathcal{G}'= \left(\mathcal{V}_{\mathcal{G}'}, \mathcal{A}_{\mathcal{G}'}\right)$ is an arc subgraph of $\mathcal{G}'' = \left(\mathcal{V}_{\mathcal{G}''}, \mathcal{A}_{\mathcal{G}''}\right)$ if $\mathcal{A}_{\mathcal{G}'} \subseteq \mathcal{A}_{\mathcal{G}''}$ and $\mathcal{V}_{\mathcal{G}'}=\mathcal{V}_{\mathcal{G}''}$. The union of $\mathcal{G}'$ and $\mathcal{G}''$ is equal to $\mathcal{G}' \cup \mathcal{G}''=\left(\mathcal{V}_{\mathcal{G}'}\cup \mathcal{V}_{\mathcal{G}''}, \mathcal{A}_{\mathcal{G}'}\cup \mathcal{A}_{\mathcal{G}''} \right)$.} of $\mathcal{G}_1$ (denoted by $\mathcal{G}_{1j}$) and an arc subgraph of $\mathcal{G}_2$ (denoted by $\mathcal{G}_{2j}$). The arc subgraphs of  $\mathcal{G}_1$ are considered as group leaders; Fig. \ref{GroupLeaders} depicts all the group leaders. For instance, $\mathcal{G}_{13}$ in this figure is the leader of group 3.  Group $j$ is the set of side information graphs constructed by the union of $\mathcal{G}_{1j}$ with each of $\{\mathcal{G}_{2k}\}_{k=1}^8$.
For instance, Fig. \ref{ElementsofGroups6} depicts the elements of group 6.

\section{Transmission Schemes}\label{transmissionschemessections}
In this section, we first establish the capacity region of six groups, stated as Theorem \ref{maintheorem}. We then enlarge the best existing inner bound for the other two groups using a joint decoding approach. Lastly, we demonstrate the looseness of the best existing inner and outer bounds.

\subsection{Deriving the Capacity for Groups 1, 2, 3, 5, 6, and 8} \label{section:capacity}

Before presenting Theorem \ref{maintheorem}, we explain our proposed capacity-achieving transmission schemes. Table \ref{TransmissionSchemes} shows these schemes for six groups. 
All the members of each group use the same scheme; there is one exception in group 5 and one in group 8 that use different schemes from other group members. 

If the codebook of the transmission scheme is composed of multiple subcodebooks, the transmitted codeword, $x^{(n)}$, is constructed from the linear superposition of multiple codewords, $\sum_k x_k^{(n)}$. Each subcodebook consists of i.i.d. codewords, $x_k^{(n)}$, generated according to an independent normal distribution $X_k \sim \mathcal{N}(0,\alpha_k P)$, where $\alpha_k\hspace{-2pt}\geq\hspace{-2pt}0$ and $\sum_k \alpha_k=1$ to satisfy the transmission power constraint. Multiplexing coding~\cite{MultiplexedCoding}, index coding~\cite{Index Coding} and dirty paper coding \cite{DPC} are employed to construct the subcodebooks. 

\begin{table*}[t]
\begin{footnotesize}
\caption{The capacity and our proposed capacity-achieving transmission scheme for different groups}
\vspace{-12pt}
\begin{center}
{\renewcommand{\arraystretch}{1.77}
\begin{tabular}{|l||l|l|}
\hline
\hspace{-5pt}Group & Transmitted Codeword & Capacity Region\\
\hline\hline
\hspace{-5pt}Group 1\hspace{-5pt} &\hspace{-4pt}$\color{blue}x_1^{(n)}\left(m_1\right)+x_2^{(n)}\left(m_2\right)+x_3^{(n)}\left(m_3\right)$
&\hspace{-4pt}\color{blue}{$R_1 < C\left(\frac{\alpha_1P}{N_1}\right)$, $R_2<C\left(\frac{\alpha_2P}{\alpha_1P+N_2}\right)$, $R_3<C\left(\frac{\alpha_3P}{(\alpha_1+\alpha_2)P+N_3}\right)$}\\
\hline
\hspace{-5pt}Group 2\hspace{-5pt} &\hspace{-4pt}$\color{blue}x_1^{(n)}\left(x_2^{(n)}\left([m_2,m_3]\right), m_1\right)+x_2^{(n)}\left([m_2,m_3]\right)$\hspace{-5pt}
&\hspace{-4pt}\color{blue}{$R_1 < C\left(\frac{\alpha_1P}{N_1}\right)$, $\sum_{i\in \{2,3\}\setminus\mathcal{O}_2}R_i<C\left(\frac{\alpha_2P}{\alpha_1P+N_2}\right)$, $R_3<C\left(\frac{\alpha_2P}{\alpha_1P+N_3}\right)$}\\
\hline
\hspace{-5pt}Group 3\hspace{-5pt}  &$\hspace{-4pt}\color{blue}x_1^{(n)}\left([m_1,m_2]\right)+x_2^{(n)}\left(m_3\right)$
&\hspace{-4pt}\color{blue}{${\sum_{i\in\{1,2\}\setminus\mathcal{O}_1}} R_i < C\left(\frac{\alpha_1 P}{N_1}\right)$, $R_2 < C\left(\frac{\alpha_1 P}{N_2}\right)$, $R_3 < C\left(\frac{\alpha_2 P}{\alpha_1 P + N_3}\right)$}\\
\hline
\hspace{-5pt}Group 5\hspace{-5pt} &\hspace{-4pt}$\color{blue}x_1^{(n)}\left([m_1,m_2]\right)+x_2^{(n)}\left([m_2,m_3]\right)$
&\hspace{-4pt}\color{blue}{${\sum_{i\notin \mathcal{O}_1}} R_i < C\left(\frac{P}{N_1}\right)$, $R_1 < C\left(\frac{\alpha_1 P}{N_1}\right)$, ${\sum_{i\notin \mathcal{O}_2}} R_i  < C\left(\frac{P}{N_2}\right)$, $R_3<C\left(\frac{\alpha_2 P}{\alpha_1 P + N_3}\right)$}\\
&\hspace{-4pt}$\mathcal{G}_{15} \cup \mathcal{G}_{22}$: &\hspace{-4pt}$\mathcal{G}_{15} \cup \mathcal{G}_{22}$:\\
&\hspace{-4pt}$\color{blue}x_1^{(n)}\left([m_1,m_2\oplus m_3]\right)+x_2^{(n)}\left(m_2\oplus m_3\right)$
&\hspace{-4pt}\color{blue}{$R_1+\max\{R_2,R_3\}< C\left(\frac{P}{N_1}\right)$, $R_1 < C\left(\frac{\alpha_1 P}{N_1}\right)$, $R_2  < C\left(\frac{P}{N_2}\right)$, $R_3<C\left(\frac{\alpha_2P}{\alpha_1 P + N_3}\right)\hspace{-6pt}$}\\
\hline
\hspace{-5pt}Group 6\hspace{-5pt}  &\hspace{-4pt}$\color{blue}x_1^{(n)}\left([m_1,m_2]\right)+x_2^{(n)}\left([m_1,m_3]\right)$
&\hspace{-4pt}\color{blue}{${\sum_{i\notin \mathcal{O}_1}} R_i < C\left(\frac{P}{N_1}\right)$, $R_2 < C\left(\frac{\alpha_1 P}{N_2}\right)$, $R_3 < C\left(\frac{\alpha_2 P}{\alpha_1 P + N_3}\right)$} \\ 
\hline
\hspace{-5pt}Group 8\hspace{-5pt}
&\hspace{-4pt}$\color{blue}x^{(n)}\left([m_1,m_2,m_3]\right)$
&\hspace{-4pt}\color{blue}{$\sum_{i\notin \mathcal{O}_1} R_i < C\left(\frac{P}{N_1}\right)$, $\sum_{i\notin \mathcal{O}_2} R_i < C\left(\frac{P}{N_2}\right)$, $R_3 < C\left(\frac{P}{N_3}\right)$}\\
&\hspace{-4pt}$\mathcal{G}_{18} \cup \mathcal{G}_{22}$: $\color{blue}x^{(n)}\left([m_1,m_2\oplus m_3]\right)$ &\hspace{-4pt}$\mathcal{G}_{18} \cup \mathcal{G}_{22}$:\color{blue}{$\;R_1+\max \{R_2,R_3\} < C\left(\frac{P}{N_1}\right)$, $R_2 < C\left(\frac{P}{N_2}\right)$, $R_3 < C\left(\frac{P}{N_3}\right)$}\\
\hline
\end{tabular}}
\label{TransmissionSchemes}
\end{center}
\end{footnotesize}
\vspace{-12pt}
\end{table*}

In multiplexing coding, two or more messages are bijectively mapped  to a single message, and then, the codewords are generated for this message. For instance, the first subcodebook of group 3 is constructed using multiplexing coding. In this scheme, the single message $M_\text{m}=[M_1,M_2]$, where $[\cdot]$ denotes the bijective map, is first formed from $M_1$ and $M_2$. Then, the codewords of the first subcodebook are generated for this single message, $M_\text{m}$, where $m_\text{m}\in \{1,2,\dotsc,2^{n(R_1+R_2)}\}$. 

In index coding (which is also called network coding \cite{NetworkCoding} in some of the works on broadcast channels), the transmitter XORs the messages to accomplish compression prior to channel coding. 
The same function can also be achieved using modulo addition~\cite{BCwithSI2UsersKramer}. 
The transmission schemes of the exceptions in groups 5 and 8  use index coding. In these schemes, $M_2\oplus M_3$ is first formed, where $\oplus$ denotes the bitwise XOR with zero padding for messages of unequal length i.e. $m_2\oplus m_3\in \{1,2,\dotsc,2^{n\max\{R_2,R_3\}}\}$. Then, the messages $M_1$ and $M_2\oplus M_3$ are fed to the channel encoder (who performs multiplexing coding and superposition coding).

Dirty paper coding is employed to construct the transmission scheme of group 2. In this scheme, first, $[M_2,M_3]$ is encoded using $2^{n(R_2+R_3)}$ i.i.d. codewords, $x_2^{(n)}\left([m_2,m_3]\right)$, generated according to $X_2\sim \mathcal{N}\left(0, \alpha_2P\right)$. Then, considering $X_2^{(n)}$ as a known non-causal interference at the transmitter, $M_1$ is encoded using dirty paper coding. The auxiliary random variable in the dirty paper coding is defined as $U=X_1+\beta X_2$ where $X_1\sim \mathcal{N}\left(0, \alpha_1P\right)$ is independent of $X_2$, and $\beta=\alpha_1P/\left(\alpha_1P+N_1\right)$.

We now state the results for the six groups in Table~\ref{TransmissionSchemes}.
\begin{theorem}\label{maintheorem} The capacity region and the optimal scheme for three-receiver AWGN BCs with private messages and side information graphs not in groups 4 and 7 are shown in  Table~\ref{TransmissionSchemes}. The capacity region for each channel is the closure of the set of all rate triplets $(R_1,R_2,R_3)$, each satisfying the conditions in the respective row for some $\alpha_k \geq 0$ such that $\sum_k \alpha_k = 1$.
\end{theorem}
\begin{IEEEproof}
The proof is presented in the appendix.
\end{IEEEproof}

\subsection{Improving the Existing Inner Bound for Groups 4 and 7} \label{section:improve}

We are unable to establish the capacity region for groups 4 and 7. However, in this subsection, we improve the best known inner bound prior to this work for these two groups. The best known inner bound, which is achieved by a separate index and channel coding scheme, is the set of all rate triples $(R_1,R_2,R_3)$, each satisfying~\cite{BCwithSI3UsersPrivateMessage}
\begin{equation}\label{deterministicregion}
\sum_{i\in\mathcal{V}_\mathcal{S}}{R_i} < \underset{i\in\mathcal{V}_\mathcal{S}}{\max} A_i,
\end{equation}
for all induced acyclic subgraphs, $\mathcal{S}$, of the side information graph. In \eqref{deterministicregion}, $A_i=\sum_{k=i}^{3}{B_k}$ where $B_1=C\left(\alpha_1P/{N_1}\right)$, $B_2=C\left({\alpha_2P}/{\left(\alpha_1 P+N_2\right)}\right)$, and $B_3=C\left({\alpha_3P}/{((\alpha_1+\alpha_2)P+N_3)}\right)$ for some $\alpha_k\geq0\;k=1,2,3$ such that $\sum_{k=1}^{3}{\alpha_k}=1$. Here, $C(q) \triangleq\frac{1}{2}\log(1+q)$.

For instance, the achievable rate region for $\mathcal{G}_{17} \cup \mathcal{G}_{24}$, a member of group 7, is the set of all rate triples $(R_1,R_2,R_3)$, each satisfying
\begin{equation}\label{region1}
\begin{split}
R_1+R_2&<B_1+B_2+B_3,\\
R_2+R_3&<B_2+B_3,\\
R_3&<B_3,
\end{split}
\end{equation}
for some $\alpha_k\geq0\;k=1,2,3$ such that $\sum_{k=1}^{3}{\alpha_k}=1$. The region in \eqref{region1} is achieved using the encoding scheme (which utilizes rate splitting, index coding, multiplexing coding and superposition coding)
\begin{align*}
x_1^{(n)}(m_{10})+x_2^{(n)}([m_{11}, m_{20}])+x_3^{(n)}([m_{21},m_{12}\oplus m_3]),
\end{align*}
and a separate index and channel decoding scheme (where side information is not utilized during channel decoding).
Using rate splitting, the message $M_1$ is divided into independent messages $M_{10}$ at rate $R_{10}$, $M_{11}$ at rate $R_{11}$, and $M_{12}$ at rate $R_{12}$ such that  $R_1=\sum_{k=0}^{2}{R_{1k}}$; the message $M_2$ is also divided into independent messages $M_{20}$ at rate $R_{20}$, and $M_{21}$ at rate $R_{21}$ such that $R_2=R_{20}+R_{21}$.
We can verify the achievability of the region in \eqref{region1} using Fourier-Motzkin elimination subsequent to successive decoding

We now show that using the same encoding scheme, but utilizing the side information during successive decoding (i.e., joint decoding), the achievable rate region can be improved. 
For the given example ($\mathcal{G}_{17} \cup \mathcal{G}_{24}$), consider the decoding of $x_3^{(n)}$ by the receivers while treating $x_1^{(n)}+x_2^{(n)}$ as noise. 
Using separate decoding, we get the condition $R_{21}+\max\{R_{12},R_3\} <B_3$ on achievability.
Using joint decoding, we can relax this condition  to $R_3 <C(\alpha_3P/N_3)$ and $R_{21}+\max\{R_{12},R_3\} <B'_3$ where $B'_3 = C\left({\alpha_3P}/{((\alpha_1+\alpha_2)P+N_2)}\right) \geq B_3$ for any choice of $\{\alpha_k\}_{k=1}^3$.
This gives an improved achievable rate region as the set of all rate triples $(R_1,R_2,R_3)$, each satisfying
\begin{equation}\label{region2}
\begin{split}
R_1+R_2&<B_1+B_2+B'_3,\\
R_2+R_3&<B_2+B'_3,\\
R_3&<\min\{C\left({\alpha_3P}/{N_3}\right),B'_3\},
\end{split}
\end{equation}
for some $\alpha_k\geq0\;k=1,2,3$ such that $\sum_{k=1}^{3}{\alpha_k}=1$. 

This joint decoding approach can be used for all the channels in groups 4 and 7 to enlarge the rate region in \eqref{deterministicregion}. However, the expression for the enlarged region depends on the particular side information configuration.

\subsection{Demonstrating the Looseness of the Existing Inner and Outer Bounds for Groups 1, 2, 3, 5, 6, and 8}

In this subsection, we demonstrate the looseness of the best known inner and outer bounds prior to this work \cite{BCwithSI3UsersPrivateMessage} compared to the capacity region of the six groups in Secion~\ref{section:capacity}.
For these six groups except group 1\footnote{For group 1, the capacity region is the same as AWGN BCs without receiver message side information.}, we can use the same argument in Section~\ref{section:improve} to show that the best known inner bound is loose. Then, the capacity region of these groups, established by our proposed transmission schemes in Secion~\ref{section:capacity}, must also be larger than the best known inner bound.

The best known outer bound states that if the rate triple $(R_1,R_2,R_3)$ is achievable, it must satisfy~\cite{BCwithSI3UsersPrivateMessage}
\begin{equation}\label{upperbound}
\sum_{i\in\mathcal{V}_\mathcal{S}}{R_i} < \underset{i\in\mathcal{V}_\mathcal{S}}{\max}\;C\left(\frac{P}{N_i}\right),
\end{equation}
for all induced acyclic subgraphs, $\mathcal{S}$, of the side information graph. The outer bound in \eqref{upperbound} is a polyhedron, and is loose for the six groups with known capacity except group 8, since the capacity-achieving transmission schemes of these groups are functions of $\alpha_k$, and therefore the capacity region has some curved surfaces. 

\begin{figure}[t]
\hspace{-25pt}
\includegraphics[width=0.58\textwidth]{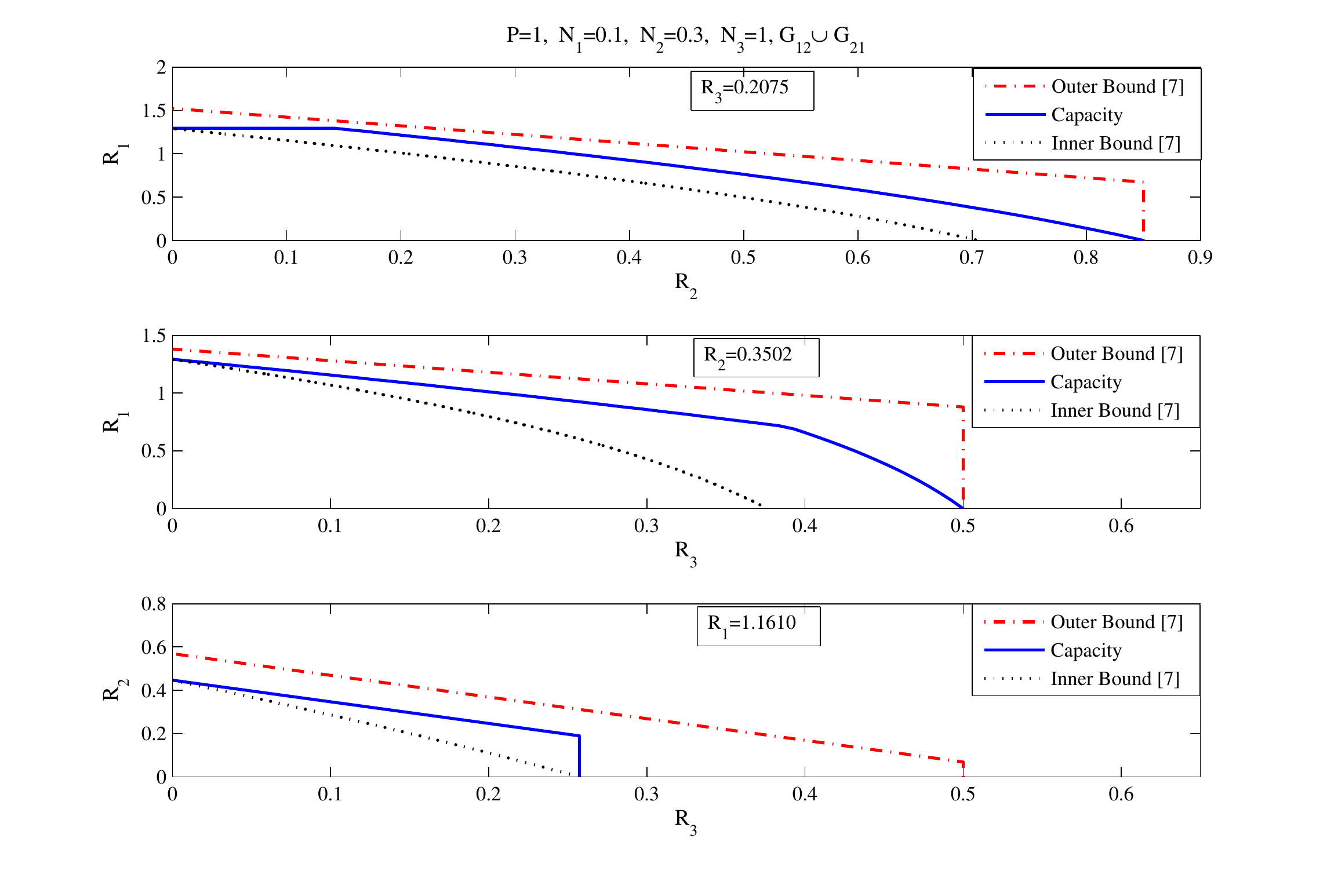}
\vspace{-34pt}
\caption{Capacity region, inner bound and outer bound comparison for $\mathcal{G}_{12}\cup\mathcal{G}_{21}$} 
\vspace{-10pt}
\label{Comparison}
\end{figure}

As an example, for $\mathcal{G}_{12}\hspace{-1pt}\cup\hspace{-1pt}\mathcal{G}_{21}$, Fig. \ref{Comparison} depicts the looseness of the best known inner and outer bounds; for this channel, the outer bound is characterized by the inequalities $R_1+R_2+R_3\hspace{-1pt}<\hspace{-1pt}C\left(P/N_1\right)$, $R_2+R_3<C\left({P}/{N_2}\right)$ and $R_3\hspace{-1pt}<\hspace{-1pt}C\left({P}/{N_3}\right)$.  
\section{Conclusion}
In this work, we have classified three-receiver AWGN BCs where the receivers have private-message requests and know the messages demanded by some other receivers as side information. The classification generates eight groups. For six groups, we have established the capacity region by proposing their capacity-achieving transmission schemes. This result (i) demonstrates the effectiveness of the classification method in building the groups with the same capacity-achieving transmission scheme and (ii) shows the looseness of the best known inner and outer bounds prior to this work. 
For the remaining two groups, we have improved the achievable rate region prior to this work by joint decoding, which utilizes side information during channel decoding.

\section*{Appendix}
In this section, we prove Theorem \ref{maintheorem}. In order to prove this theorem, we present the achievability and the converse proofs for the six groups in Table \ref{TransmissionSchemes}. The proofs are based on those for AWGN BCs without side information \cite{AWGNBCConverse, NITBook}. In the converse, we use Fano's inequality and the entropy power inequality (EPI). Based on Fano's inequality,
\begin{equation}
H(M_i \mid Y_i^{(n)},\mathcal{K}_i)\leq n\epsilon_{n,i},\quad i = 1,2,3\label{fano},
\end{equation}
where $\epsilon_{n,i} \rightarrow 0$ as $n\rightarrow \infty$. For the sake of simplicity we use $\epsilon_n$ instead of $\epsilon_{n,i}$ for the remainder. In the converse, we also use the fact that the capacity region of a stochastically degraded broadcast channel without feedback is the same as its equivalent physically degraded broadcast channel \cite[p. 444]{NITBook} where the channel input and outputs form a Markov chain, $X\rightarrow Y_{1}\rightarrow Y_{2} \rightarrow  Y_{3}$, i.e., 
\begin{align}\label{physicallydegardedness}
&Y_1=X+Z_1,\nonumber\\
&Y_i=Y_{i-1}+\tilde{Z}_i \;\;i=2,3,
\end{align}
where $\tilde{Z}_i \sim \mathcal{N}\left(0, N_i-N_{i-1}\right)\;i=2,3$. 

Before presenting the proof of Theorem \ref{maintheorem}, we prove some lemmas that will be used in the converse.
\begin{lemma} \label{modifiedfano} If $\mathcal{L}\subseteq\left\{M_1,M_2,M_3\right\}$, then 
\vspace{-4pt}
\begin{multline*}
H\left(M_l \mid Y_i^n,\mathcal{L}\right) \leq H\left(M_l \mid Y_j^n,\mathcal{L}\right),\\ \forall l,i,j \in \{1,2,3\} \text{ such that } i < j. 
\end{multline*}
\end{lemma}
\vspace{-4pt}
\begin{IEEEproof}
The proof is similar to the proof for the data processing inequality \cite[p. 25]{NITBook}. We just need to expand $I\hspace{-2pt}\left(M_l;Y_i^{(n)}\hspace{-2pt},Y_j^{(n)}\hspace{-2pt}\mid\hspace{-2pt}\mathcal{L}\right)$ in two ways by the mutual information chain rule and use the Markov chain, resulted from the physically degradedness. Using the mutual information chain rule, we have
\begin{align}\label{chainrule1}
&I\left(M_l;Y_i^{(n)},Y_j^{(n)},\mathcal{L}\right) \nonumber \\
&=I\left(M_l;Y_i^{(n)},\mathcal{L}\right)+I\left(M_l;Y_j^{(n)} \mid Y_i^{(n)},\mathcal{L}\right) \nonumber\\
&=I\left(M_l;Y_i^{(n)},\mathcal{L}\right) \nonumber\\
&\hskip20pt+h\left(Y_j^{(n)} \mid Y_i^{(n)},\mathcal{L}\right)-h\left(Y_j^{(n)} \mid Y_i^{(n)},\mathcal{L},M_l\right) \nonumber\\
&\overset{(a)}{=}I\left(M_l;Y_i^{(n)},\mathcal{L}\right)+h\left(Y_j^{(n)} \mid Y_i^{(n)}\right)-h\left(Y_j^{(n)} \mid Y_i^{(n)}\right) \nonumber\\
&=I\left(M_l;Y_i^{(n)},\mathcal{L}\right),
\end{align}
where $(a)$ follows from the physically degradedness of the channel. Using the mutual information chain rule again, we have
\begin{multline}\label{chainrule2}
I\left(M_l;Y_i^{(n)},Y_j^{(n)},\mathcal{L}\right)\\
=I\left(M_l;Y_j^{(n)},\mathcal{L}\right)+I\left(M_l;Y_i^{(n)} \mid Y_j^{(n)},\mathcal{L}\right).
\end{multline}
Since (\ref{chainrule1}) and (\ref{chainrule2}) are equal, we have
\begin{align}\label{mutualinequality}
I\left(M_l;Y_i^{(n)},\mathcal{L}\right) \geq I\left(M_l;Y_j^{(n)},\mathcal{L}\right).
\end{align}
By replacing both sides of the inequality in (\ref{mutualinequality}) with the following equations, the proof is complete.
\begin{align*}
&I\left(M_l;Y_i^{(n)},\mathcal{L}\right)=H\left(M_l\right)-H\left(M_l \mid Y_i^{(n)},\mathcal{L}\right),\\
&I\left(M_l;Y_j^{(n)},\mathcal{L}\right)=H\left(M_l\right)-H\left(M_l \mid Y_j^{(n)},\mathcal{L}\right).
\end{align*}
\end{IEEEproof}
\begin{lemma}[EPI inequalities]\label{epiinequalities} If $\mathcal{L} \subseteq \left\{M_1,M_2,M_3\right\}$ and $h(Y_2^{(n)} \mid \mathcal{L})=\frac{n}{2} \log {2\pi e\left(\alpha P+N_2\right)}$ for an $0\leq \alpha \leq1$ then
\begin{align}
h\left(Y_1^{(n)} \mid \mathcal{L}\right) \leq \frac{n}{2} \log {2\pi e\left(\alpha P+N_1\right)}, \label{epi1}\\
h\left(Y_3^{(n)} \mid \mathcal{L}\right) \geq \frac{n}{2} \log {2\pi e\left(\alpha P+N_3\right)}. \label{epi2}
\end{align}
\end{lemma}
\begin{IEEEproof}
Based on the conditional EPI {\cite[p. 22]{NITBook}} and \eqref{physicallydegardedness}, we have
\begin{align}
2^{\frac{2}{n}h\left(Y_2^{(n)} \mid \mathcal{L} \right)} \geq 2^{\frac{2}{n}h\left(Y_1^{(n)} \mid \mathcal{L} \right)}+2^{\frac{2}{n}h\left(\tilde{Z}_2^{(n)}\mid \mathcal{L} \right)}, \label{cepi1}\\
2^{\frac{2}{n}h\left(Y_3^{(n)}\mid \mathcal{L} \right)} \geq 2^{\frac{2}{n}h\left(Y_2^{(n)}\mid \mathcal{L} \right)}+2^{\frac{2}{n}h\left(\tilde{Z}_3^{(n)}\mid \mathcal{L} \right)}. \label{cepi2}
\end{align}
Using $h(Y_2^{(n)} \mid \mathcal{L})=\frac{n}{2} \log {2\pi e\left(\alpha P+N_2\right)}$ and the following equations in (\ref{cepi1}) and (\ref{cepi2}) completes the proof.
\begin{align*}
h(\tilde{Z}_2^{(n)}\mid \mathcal{L})=h(\tilde{Z}_2^{(n)})=\frac{n}{2} \log 2 \pi e\left(N_2-N_1\right),\\
h(\tilde{Z}_3^{(n)}\mid \mathcal{L})=h(\tilde{Z}_3^{(n)})=\frac{n}{2} \log 2 \pi e\left(N_3-N_2\right). 
\end{align*}
\end{IEEEproof}
\begin{corollary}\label{epicoro} If $\mathcal{L} \subseteq \left\{M_1,M_2,M_3\right\}$ and $h(Y_3^{(n)} \mid \mathcal{L})=\frac{n}{2} \log {2\pi e\left(\alpha P+N_3\right)}$ for an $0\leq \alpha \leq1$, then
\begin{align}
h\left(Y_1^{(n)} \mid \mathcal{L}\right) \leq \frac{n}{2} \log {2\pi e\left(\alpha P+N_1\right)}, \label{epi1}\\
h\left(Y_2^{(n)} \mid \mathcal{L}\right) \leq \frac{n}{2} \log {2\pi e\left(\alpha P+N_2\right)}. \label{epi2}
\end{align}
\end{corollary}

We now present the proof of Theorem \ref{maintheorem}.
\begin{IEEEproof} As mentioned earlier, in order to prove this theorem, we present the achievability and the converse proofs for the six groups in Table \ref{TransmissionSchemes}.

Group 1: The achievability of the given rate region for this group is proved by using successive decoding at the receivers where each receiver decodes the requested message of the weakest receiver (if it is unknown), followed by that of the next weaker receiver, and so on until its own requested message. To prove the converse for the members of this group, the given rate region for this group is reformulated as
\begin{align*}
& R_1 < \frac{1}{2}\log\left(\frac{\beta_1 P+N_1}{\beta_0P+N_1}\right),\\
&R_2 < \frac{1}{2}\log\left(\frac{\beta_2 P+N_2}{\beta_1P+N_2}\right),\\
&R_3 < \frac{1}{2}\log\left(\frac{\beta_3 P+N_3}{\beta_2P+N_3}\right),
\end{align*}
where $0=\beta_0\leq \beta_1\leq \beta_2 \leq \beta_3=1$, $\beta_1=\alpha_1$ and $\beta_2=\alpha_1+\alpha_2$. The converse for the group leader (the member without side information) was proved by Bergmans \cite{AWGNBCConverse}. To prove the converse for the other members, we only need to prove the converse for the member with maximum possible side information, $\mathcal{G}_{11}\cup\mathcal{G}_{28}$, where $\mathcal{K}_1= \{M_2,M_3\}$, $\mathcal{K}_2=\{M_3\}$ and $\mathcal{K}_3=\varnothing$. This is because the capacity region of this member can be considered as an outer bound to the capacity region of the other members, and the capacity region of the group leader is an inner bound to the capacity region of the others. Then, proving the converse for $\mathcal{G}_{11}\cup\mathcal{G}_{28}$ shows that the inner and outer bounds coincide and the proof for this group is complete. 

Here, we prove the converse for $\mathcal{G}_{11}\cup\mathcal{G}_{28}$. In this channel, $R_3$ is upper bounded as
\begin{align}\label{proof11}
&nR_3=H(M_3)=H(M_3 \mid Y_3^{(n)})+I(Y_3^{(n)};M_3) \nonumber\\
&=H(M_3 \mid Y_3^{(n)})+h(Y_3^{(n)})-h(Y_3^{(n)}\mid M_3) \nonumber\\
&\overset{(a)}{\leq} n\epsilon_n+h(Y_3^{(n)})-h(Y_3^{(n)}\mid M_3) \nonumber\\
&\overset{(b)}{\leq} n\epsilon_n+\frac{n}{2}\log2\pi e(P+N_3)-h(Y_3^{(n)}\mid M_3) \nonumber\\
&\overset{(c)}{=}n\epsilon_n+\frac{n}{2}\log2\pi e(P+N_3)-\frac{n}{2}\log2\pi e(\beta_2 P+N_3),
\end{align}
where (a) follows from \eqref{fano}, (b) from $h(Y_3^{(n)})\leq\frac{n}{2}\log2\pi e(P+N_3)$, and (c) from the fact that
\begin{multline*}
\frac{n}{2}\log2\pi e N_3=h(Z_3^{(n)})= h(Y_3^{(n)}\mid X^{(n)})\\
\overset{(d)}{\leq} h(Y_3^{(n)}\mid M_3)\leq h(Y_3^{(n)}) \leq \frac{n}{2}\log2\pi e(P+N_3),
\end{multline*}
where (d) is because $M_3\rightarrow X^{(n)}\rightarrow Y_3^{(n)}$ form a Markov chain; then since $\frac{n}{2}\log2\pi e N_3\leq h(Y_3^{(n)}\mid M_3)\leq \frac{n}{2}\log2\pi e(P+N_3)$ there must exist a $0\leq\beta_2\leq1$ such that  $h(Y_3^{(n)}\mid M_3)=\frac{n}{2}\log2\pi e(\beta_2 P+N_3)$.

$R_2$ is upper bounded as
\begin{align}\label{proof12}
&nR_2=H(M_2)=H(M_2 \mid Y_2^{(n)},M_3)+I(M_2;Y_2^{(n)},M_3) \nonumber\\
&\overset{(a)}{=}H(M_2 \mid Y_2^{(n)},M_3)+I(M_2;Y_2^{(n)}\mid M_3) \nonumber\\
&=H(M_2 \mid Y_2^{(n)},M_3)+h(Y_2^{(n)}\mid M_3)-h(Y_2^{(n)}\mid M_2,M_3) \nonumber\\
&\overset{(b)}{\leq} n\epsilon_n+h(Y_2^{(n)}\mid M_3)-h(Y_2^{(n)}\mid M_2,M_3) \nonumber\\
&\overset{(c)}{\leq} n\epsilon_n+\frac{n}{2}\log2\pi e(\beta_2 P+N_2)-h(Y_2^{(n)}\mid M_2,M_3) \nonumber\\
&\overset{(d)}{=} n\epsilon_n+\frac{n}{2}\log2\pi e(\beta_2 P+N_2)-\frac{n}{2}\log2\pi e(\beta_1 P+N_2),
\end{align}
where (a) follows from the independence of $M_2$ and $M_3$, (b) from \eqref{fano}, (c) from Corollary \ref{epicoro} and $h(Y_3^{(n)}\mid M_3)=\frac{n}{2}\log2\pi e(\beta_2 P+N_3)$, and (d) is due to
\begin{multline*}
\frac{n}{2}\log2\pi e N_2=h(Z_2^{(n)})= h(Y_2^{(n)}\mid X^{(n)})\overset{(e)}{\leq}\\
\hspace{5pt}h(Y_2^{(n)}\mid M_2,M_3)\leq h(Y_2^{(n)}\mid M_3) \leq \frac{n}{2}\log2\pi e(\beta_2P+N_2),
\end{multline*}
where (e) is because $(M_2,M_3)\rightarrow X^{(n)}\rightarrow Y_2^{(n)}$ form a Markov chain; then since $\frac{n}{2}\log2\pi e N_2\leq h(Y_2^{(n)}\mid M_2,M_3)\leq \frac{n}{2}\log2\pi e(\beta_2P+N_2)$ there must exist a $0\leq\beta_1\leq \beta_2$ such that  $h(Y_2^{(n)}\mid M_2,M_3)=\frac{n}{2}\log2\pi e(\beta_1 P+N_2)$.

$R_1$ is also upper bounded as
\begin{align}\label{proof13}
&nR_1=H(M_1) \nonumber\\
&=H(M_1 \mid Y_1^{(n)},M_2,M_3)+I(M_1;Y_1^{(n)},M_2,M_3) \nonumber\\
&\overset{(a)}{=}H(M_1 \mid Y_1^{(n)},M_2,M_3)+I(M_1;Y_1^{(n)}\mid M_2,M_3) \nonumber\\
&=H(M_1 \mid Y_1^{(n)},M_2,M_3) \nonumber\\
&\hskip30pt+h(Y_1^{(n)}\mid M_2,M_3)-h(Y_1^{(n)}\mid M_1,M_2,M_3) \nonumber\\
&\overset{(b)}{\leq} n\epsilon_n+h(Y_1^{(n)}\mid M_2,M_3)-h(Y_1^{(n)}\mid M_1,M_2,M_3) \nonumber\\
&\overset{(c)}{\leq} n\epsilon_n+\frac{n}{2}\log2\pi e(\beta_1 P+N_1)-h(Y_1^{(n)}\mid M_1,M_2,M_3) \nonumber\\
&\overset{(d)}{\leq}n\epsilon_n+\frac{n}{2}\log2\pi e(\beta_1 P+N_1)-\frac{n}{2}\log2\pi e N_1,
\end{align}
where (a) follows from the independence of $M_1$, $M_2$ and $M_3$, (b) from \eqref{fano}, (c) from Lemma \ref{epiinequalities} and $h(Y_2^{(n)}\mid M_2,M_3)=\frac{n}{2}\log2\pi e(\beta_1 P+N_2)$, and (d) from 
\begin{multline}\label{noise1entropy}
h(Y_1^{(n)}\mid M_1,M_2,M_3)\\
\overset{(e)}{\geq}(Y_1^{(n)}\mid X^{(n)})=h(Z_1^{(n)})=\frac{n}{2}\log2\pi e N_1,
\end{multline}
where (e) is because $(M_1,M_2,M_3)\rightarrow X^{(n)}\rightarrow Y_1^{(n)}$ form a Markov chain.

From (\ref{proof11})--(\ref{proof13}) and since $\epsilon_n$ goes to zero as $n \rightarrow \infty$, the converse proof for $\mathcal{G}_{11}\cup\mathcal{G}_{28}$ and this group is complete.

Group 2: The achievability of the given rate region for this group is proved by considering two points during the decoding. First, receivers 2 and 3 consider $x_1^{(n)}$ as noise and decode $x_2^{(n)}$ based on their side information. Second, receiver 1, benefiting from dirty paper coding, is not affected by $x_2^{(n)}$ irrespective of its knowledge about $M_2$ and $M_3$.  

Here, we prove the converse for $\mathcal{G}_{12}\cup\mathcal{G}_{21}$. For this channel we have
\begin{align}\label{proof24}
&n(R_2+R_3)=H(M_2,M_3)\nonumber\\
&=H(M_2,M_3 \mid Y_2^{(n)})+I(M_2,M_3;Y_2^{(n)}) \nonumber\\
&=H(M_2,M_3 \mid Y_2^{(n)})+h(Y_2^{(n)})-h(Y_2^{(n)}\mid M_2,M_3) \nonumber\\
&\overset{(a)}{\leq}2 n\epsilon_n+h(Y_2^{(n)})-h(Y_2^{(n)}\mid M_2,M_3) \nonumber\\
&\overset{(b)}{\leq}2 n\epsilon_n+\frac{n}{2}\log2\pi e(P+N_2)-h(Y_2^{(n)}\mid M_2,M_3) \nonumber\\
&\overset{(c)}{=}2 n\epsilon_n+\frac{n}{2}\log2\pi e(P+N_2)-\frac{n}{2}\log2\pi e(\alpha P+N_2),
\end{align}
where (a) follows from adding the following inequalities which are the results of using Lemma \ref{modifiedfano} and \eqref{fano} as
\begin{align*}
&H(M_3\mid Y_2^{(n)},M_2)\leq H(M_3\mid Y_3^{(n)},M_2) \leq n\epsilon_n,\\
&H(M_2\mid Y_2^{(n)})\leq n \epsilon_n.
\end{align*}
In \eqref{proof24}, (b) follows from $h(Y_2^{(n)}) \leq \frac{n}{2}\log2\pi e(P+N_2)$ and (c) from
\begin{multline*}
\frac{n}{2}\log2\pi e N_2=h(Z_2^{(n)})= h(Y_2^{(n)}\mid X^{(n)})\\
\overset{(d)}{\leq} h(Y_2^{(n)}\mid M_2,M_3)\leq h(Y_2^{(n)}) \leq \frac{n}{2}\log2\pi e(P+N_2),
\end{multline*}
where (d) is because $(M_2,M_3)\rightarrow X^{(n)}\rightarrow Y_2^{(n)}$ form a Markov chain; then since $\frac{n}{2}\log2\pi e N_2\leq h(Y_2^{(n)}\mid M_2,M_3)\leq \frac{n}{2}\log2\pi e(P+N_2)$ 
there must exist an $0\leq \alpha \leq1$ such that $h(Y_2^{(n)}\mid M_2,M_3)=\frac{n}{2}\log2\pi e(\alpha P+N_2)$.

In this channel, $R_3$ is upper bounded as
\begin{align}\label{proof25}
&nR_3=H(M_3)=H(M_3 \mid Y_3^{(n)},M_2)+I(M_3;Y_3^{(n)},M_2) \nonumber\\
&\overset{(a)}{=}H(M_3 \mid Y_3^{(n)},M_2)+I(M_3;Y_3^{(n)}\mid M_2) \nonumber\\
&=H(M_3 \mid Y_3^{(n)},M_2)+h(Y_3^{(n)}\mid M_2)-h(Y_3^{(n)}\mid M_2,M_3) \nonumber\\
&\overset{(b)}{\leq} n\epsilon_n+h(Y_3^{(n)}\mid M_2)-h(Y_3^{(n)}\mid M_2,M_3) \nonumber\\
&\overset{(c)}{\leq} n\epsilon_n+\frac{n}{2}\log2\pi e(P+N_3)-h(Y_3^{(n)}\mid M_2,M_3) \nonumber\\
&\overset{(d)}{\leq}n\epsilon_n+\frac{n}{2}\log2\pi e(P+N_3)-\frac{n}{2}\log2\pi e(\alpha P+N_3),
\end{align}
where (a) follows from the independence of $M_2$ and $M_3$, (b) from \eqref{fano}, (c) from $h(Y_3^{(n)}\mid M_2)\leq h(Y_3^{(n)}) \leq\frac{n}{2}\log2\pi e(P+N_3)$, and (d) from Lemma \ref{epiinequalities} and $h(Y_2^{(n)}\mid M_2,M_3)=\frac{n}{2}\log2\pi e(\alpha P+N_2)$.  

In this channel, $R_1$ is also upper bounded as
\begin{align}\label{proof26}
&nR_1=H(M_1)\nonumber\\
&=H(M_1 \mid Y_1^{(n)},M_2,M_3)+I(M_1;Y_1^{(n)},M_2,M_3) \nonumber\\
&\overset{(a)}{=}H(M_1 \mid Y_1^{(n)},M_2,M_3)+I(M_1;Y_1^{(n)} \mid M_2,M_3) \nonumber\\
&=H(M_1 \mid Y_1^{(n)},M_2,M_3)\nonumber\\
&\hskip50pt+h(Y_1^{(n)}\mid M_2,M_3)-h(Y_1^{(n)}\mid M_1,M_2,M_3) \nonumber\\
&\overset{(b)}{\leq} n\epsilon_n+h(Y_1^{(n)}\mid M_2,M_3)-h(Y_1^{(n)}\mid M_1,M_2,M_3) \nonumber\\
&\overset{(c)}{\leq} n\epsilon_n+\frac{n}{2}\log2\pi e(\alpha P+N_1)-h(Y_1^{(n)}\mid M_1,M_2,M_3) \nonumber\\
&\overset{(d)}{\leq}n\epsilon_n+\frac{n}{2}\log2\pi e(\alpha P+N_1)-\frac{n}{2}\log2\pi e N_1,
\end{align}
where (a) follows from the independence of $M_1$, $M_2$ and $M_3$, (b) from \eqref{fano} and $H(M_1 \mid Y_1^{(n)},M_2,M_3)\leq H(M_1 \mid Y_1^{(n)})$, (c) from Lemma \ref{epiinequalities} and $h(Y_2^{(n)}\mid M_2,M_3)=\frac{n}{2}\log2\pi e(\alpha P+N_2)$, and (d) from (\ref{noise1entropy}).

From (\ref{proof24})--(\ref{proof26}) and since $\epsilon_n$ goes to zero as $n \rightarrow \infty$, the converse proof for this member is complete. 

The inequalities (\ref{proof24}), (\ref{proof25}), and (\ref{proof26}) also hold for all other members in this group, but if receiver 2 knows $M_3$, we need to modify \eqref{proof24} as follows to prove the converse.
\begin{align}\label{proof21} 
&nR_2=H(M_2\nonumber)\\
&=H(M_2 \mid Y_2^{(n)},M_3)+I(M_2;Y_2^{(n)},M_3) \nonumber\\
&\overset{(a)}{=}H(M_2 \mid Y_2^{(n)},M_3)+I(M_2;Y_2^{(n)} \mid M_3) \nonumber\\
&=H(M_2 \mid Y_2^{(n)},M_3)+h(Y_2^{(n)}\mid M_3)-h(Y_2^{(n)}\mid M_2,M_3) \nonumber\\
&\overset{(b)}{\leq}n\epsilon_n+h(Y_2^{(n)}\mid M_3)-h(Y_2^{(n)}\mid M_2,M_3) \nonumber\\
&\overset{(c)}{\leq}n\epsilon_n+\frac{n}{2}\log2\pi e(P+N_2)-h(Y_2^{(n)}\mid M_2,M_3) \nonumber\\
&\overset{(d)}{=}n\epsilon_n+\frac{n}{2}\log2\pi e(P+N_2)-\frac{n}{2}\log2\pi e(\alpha P+N_2),
\end{align}
where (a) follows from the independence of $M_3$ and $M_2$, (b) from \eqref{fano}, (c) from $h(Y_2^{(n)}\mid M_3)\leq h(Y_2^{(n)})\leq \frac{n}{2}\log2\pi e(P+N_2)$, and (d) from
\begin{multline*}
\frac{n}{2}\log2\pi e N_2=h(Z_2^{(n)})= h(Y_2^{(n)}\mid X^{(n)})\\
\overset{(e)}{\leq} h(Y_2^{(n)}\mid M_2,M_3)\leq h(Y_2^{(n)}) \leq \frac{n}{2}\log2\pi e(P+N_2),
\end{multline*}
where (e) is because $(M_2,M_3)\rightarrow X^{(n)}\rightarrow Y_2^{(n)}$ form a Markov chain; then since $\frac{n}{2}\log2\pi e N_2\leq h(Y_2^{(n)}\mid M_2,M_3)\leq \frac{n}{2}\log2\pi e(P+N_2)$ 
there must exist an $0\leq \alpha \leq1$ such that $h(Y_2^{(n)}\mid M_2,M_3)=\frac{n}{2}\log2\pi e(\alpha P+N_2)$.

Group 3: The achievability of the given rate region for this group is proved via successive decoding. Receiver 3 considers $x_1^{(n)}$ as noise and decodes $x_2^{(n)}$; Receivers 1 and 2 first decode $x_2^{(n)}$ (if $M_3$ is unknown to them) while treating $x_1^{(n)}$ as noise and then decode $x_1^{(n)}$.

The converse proof for $\mathcal{G}_{13}\cup\mathcal{G}_{21}$ is as follows. In this channel, $R_3$ is upper bounded as
\begin{align}\label{proof34}
&nR_3=H(M_3)=H(M_3 \mid Y_3^{(n)})+I(M_3;Y_3^{(n)}) \nonumber\\
&=H(M_3 \mid Y_3^{(n)})+h(Y_3^{(n)})-h(Y_3^{(n)}\mid M_3) \nonumber\\
&\overset{(a)}{\leq} n\epsilon_n+h(Y_3^{(n)})-h(Y_3^{(n)}\mid M_3) \nonumber\\
&\overset{(b)}{\leq} n\epsilon_n+\frac{n}{2}\log2\pi e(P+N_3)-h(Y_3^{(n)}\mid M_3) \nonumber\\
&\overset{(c)}{=}n\epsilon_n+\frac{n}{2}\log2\pi e(P+N_3)-\frac{n}{2}\log2\pi e(\alpha P+N_3),
\end{align}
where (a) follows from \eqref{fano}, (b) from $h(Y_3^{(n)})\leq \frac{n}{2}\log2\pi e(P+N_3)$, and (c) from the fact that
\begin{multline*}
\frac{n}{2}\log2\pi e N_3=h(Z_3^{(n)})= h(Y_3^{(n)}\mid X^{(n)})\\
\overset{(d)}{\leq} h(Y_3^{(n)}\mid M_3)\leq h(Y_3^{(n)}) \leq \frac{n}{2}\log2\pi e(P+N_3),
\end{multline*}
where (d) is because $M_3\rightarrow X^{(n)}\rightarrow Y_3^{(n)}$ form a Markov chain; then since $\frac{n}{2}\log2\pi e N_3\leq h(Y_3^{(n)}\mid M_3)\leq \frac{n}{2}\log2\pi e(P+N_3)$ there must exist an $0\leq \alpha \leq1$ such that $h(Y_3^{(n)}\mid M_3)=\frac{n}{2}\log2\pi e(\alpha P+N_3)$.

In this channel, $R_2$ is upper bounded as
\begin{align}\label{proof35}
&nR_2=H(M_2)\nonumber\\
&=H(M_2 \mid Y_2^{(n)},M_1,M_3)+I(M_2;Y_2^{(n)},M_1,M_3) \nonumber\\
&\overset{(a)}{=}H(M_2 \mid Y_2^{(n)},M_1,M_3)+I(M_2;Y_2^{(n)} \mid M_1,M_3) \nonumber\\
&=H(M_2 \mid Y_2^{(n)},M_3,M_1)\nonumber\\
&\hskip40pt+h(Y_2^{(n)}\mid M_1,M_3)-h(Y_2^{(n)}\mid M_1,M_2,M_3) \nonumber\\
&\overset{(b)}{\leq}n\epsilon_n+h(Y_2^{(n)}\mid M_1,M_3)-h(Y_2^{(n)}\mid M_1,M_2,M_3) \nonumber\\
&\overset{(c)}{\leq}n\epsilon_n+h(Y_2^{(n)}\mid M_3)-h(Y_2^{(n)}\mid M_1,M_2,M_3) \nonumber\\
&\overset{(d)}{\leq} n\epsilon_n+\frac{n}{2}\log2\pi e(\alpha P+N_2)-h(Y_2^{(n)}\mid M_1,M_2,M_3)\nonumber\\
&\overset{(e)}{\leq} n\epsilon_n+\frac{n}{2}\log2\pi e(\alpha P+N_2)-\frac{n}{2}\log2\pi e N_2,
\end{align}
where (a) follows from the independence of $M_1$, $M_2$ and $M_3$, (b) from \eqref{fano} and $H(M_2 \mid Y_2^{(n)},M_1,M_3)\leq H(M_2 \mid Y_2^{(n)},M_1)$, (c) from $h(Y_2^{(n)}\mid M_1,M_3)\leq h(Y_2^{(n)}\mid M_3)$, (d) from Corollary \ref{epicoro} and $h(Y_3^{(n)}\mid M_3)=\frac{n}{2}\log2\pi e(\alpha P+N_3)$, and (e) from 
\begin{multline}\label{noise2entropy}
h(Y_2^{(n)}\mid M_1,M_2,M_3)\\
\overset{(f)}{\geq}(Y_2^{(n)}\mid X^{(n)})=h(Z_2^{(n)})=\frac{n}{2}\log2\pi e N_2,
\end{multline}
where (f) is because $(M_1,M_2,M_3)\rightarrow X^{(n)}\rightarrow Y_2^{(n)}$ form a Markov chain.

For this channel, we also have
\begin{align}\label{proof36}
&n(R_1+R_2)=H(M_1,M_2)\nonumber\\
&=H(M_1, M_2 \mid Y_1^{(n)},M_3)+I(M_1, M_2;Y_1^{(n)},M_3) \nonumber\\
&\overset{(a)}{=}H(M_1, M_2 \mid Y_1^{(n)},M_3)+I(M_1, M_2;Y_1^{(n)} \mid M_3) \nonumber\\
&=H(M_1, M_2 \mid Y_1^{(n)},M_3)\nonumber\\
&\hskip40pt+h(Y_1^{(n)}\mid M_3)-h(Y_1^{(n)}\mid M_1,M_2,M_3) \nonumber\\
&\overset{(b)}{\leq}2 n\epsilon_n+h(Y_1^{(n)}\mid M_3)-h(Y_1^{(n)}\mid M_1,M_2,M_3) \nonumber\\
&\overset{(c)}{\leq} 2 n\epsilon_n+\frac{n}{2}\log2\pi e(\alpha P+N_1)-h(Y_1^{(n)}\mid M_1,M_2,M_3)\nonumber\\
&\overset{(d)}{\leq} 2 n\epsilon_n+\frac{n}{2}\log2\pi e(\alpha P+N_1)-\frac{n}{2}\log2\pi e N_1,
\end{align}
where (a) follows from the independence of $M_1$, $M_2$ and $M_3$, and (b) from adding the following inequalities which are the results of using Lemma \ref{modifiedfano} and \eqref{fano} as
\begin{align*}
&H(M_2 \mid Y_1^{(n)},M_1,M_3) \leq H(M_2 \mid Y_1^{(n)},M_1)\\
&\hskip110pt\leq H(M_2 \mid Y_2^{(n)},M_1) \leq n \epsilon_n,\\
&H(M_1 \mid Y_1^{(n)},M_3) \leq H(M_1 \mid Y_1^{(n)}) \leq n \epsilon_n.
\end{align*}
In \eqref{proof36}, (c) follows from Corollary \ref{epicoro} and $h(Y_3^{(n)}\mid M_3)=\frac{n}{2}\log2\pi e(\alpha P+N_3)$, and (d) from  (\ref{noise1entropy}).

From (\ref{proof34}), (\ref{proof35}), (\ref{proof36}) and since $\epsilon_n$ goes to zero as $n \rightarrow \infty$, the converse proof for this member is complete. 

The converse proof for all other members in this group is straightforward; we only need to modify \eqref{proof36} if receiver 1 knows $M_2$.

Group 5: The achievablity for $\mathcal{G}_{15} \cup \mathcal{G}_{21}$ is proved by using successive decoding at receiver 3, and simultaneous decoding at receivers 1 and 2. Since receivers 3 know $M_2$, $R_3< C(\alpha_2P/(\alpha_1P+N_3))$ is required for achievability concerning this receiver. Receiver 2, using simultaneous decoding, decodes $\hat{m}_2$ if there exists a unique $\hat{m}_2$ such that $(X_1^{(n)}([1,\hat{m}_2]),X_2^{(n)}([\hat{m}_2,m_3]),Y_2^{(n)})\in \mathcal{T}_\delta^{(n)}$ for some $m_3$, where $\mathcal{T}_\delta^{(n)}$ is the set of jointly $\delta$-typical $n$-sequences~\cite[p.\ 521]{ITBook}; otherwise the error is declared. Assuming the transmitted messages are equal to one by the symmetry of the code generation, the error events at receiver 2 for $\mathcal{G}_{15} \cup \mathcal{G}_{21}$ are
\begin{align*}
&\mathcal{E}_{21}: \left(X_1^{(n)}([1,1]),X_2^{(n)}([1,m_3]),Y_2^{(n)}\right)\notin \mathcal{T}_\delta^{(n)} \hspace{3pt}\text{for all}\;m_3,\\
&\mathcal{E}_{22}:\left(X_1^{(n)}([1,m_2]),X_2^{(n)}([m_2,m_3]),Y_2^{(n)}\right)\in \mathcal{T}_\delta^{(n)}\\ &\hskip155pt\mathrm{for\;some}\;m_2\neq1,m_3.
\end{align*}
From the properties of joint typicality~\cite[Theorems 15.2.1 and 15.2.3]{ITBook}, it can be seen for $\mathcal{G}_{15}\cup\mathcal{G}_{21}$, $R_2+R_3<C(P/N_2)$, guarantees that the probability of error at receiver 2 tends to zero as $n$ increases. Receiver 1, using simultaneous decoding, decodes $\hat{m}_1$ if there exists a unique $\hat{m}_1$ such that $(X_1^{(n)}([\hat{m}_1,m_2]),X_2^{(n)}([m_2,m_3]),Y_1^{(n)})\in \mathcal{T}_\delta^{(n)}$ for some $m_2,m_3$; otherwise the error is declared. The error events at receiver 1 for $\mathcal{G}_{15} \cup \mathcal{G}_{21}$ are
\begin{align*}
&\mathcal{E}_{11}:\left(X_1^{(n)}([1,m_2]),X_2^{(n)}([m_2,m_3]),Y_1^{(n)}\right)\notin \mathcal{T}_\delta^{(n)}\\ 
&\hskip180pt\text{for all}\;m_2,m_3,\\
&\mathcal{E}_{12}:\left(X_1^{(n)}([m_1,1]),X_2^{(n)}([1,1]),Y_1^{(n)}\right)\in \mathcal{T}_\delta^{(n)}\\ 
&\hskip170pt\mathrm{for\;some}\;m_1\neq1,\\
&\mathcal{E}_{13}:\left(X_1^{(n)}([m_1,1]),X_2^{(n)}([1,m_3]),Y_1^{(n)}\right)\in \mathcal{T}_\delta^{(n)}\\ 
&\hskip135pt \mathrm{for\;some}\;m_1\neq1, m_3\neq1,\\
&\mathcal{E}_{14}:\left(X_1^{(n)}([m_1,m_2]),X_2^{(n)}([m_2,m_3]),Y_1^{(n)}\right)\in \mathcal{T}_\delta^{(n)}\\ 
&\hskip118pt \mathrm{for\;some}\;m_1\neq1, m_2\neq1, m_3.
\end{align*}
According to these error events, for $\mathcal{G}_{15}\cup\mathcal{G}_{21}$, $R_1+R_2+R_3<C(P/N_1)$ and $R_1<C(\alpha_1P/N_1)$ are required for achievability concerning receiver 1. 

For all other members in group 5 except $\mathcal{G}_{15}\cup\mathcal{G}_{22}$, we use the same encoding and decoding schemes, but each receiver makes its decoding decision based on its extra side information.

Here, we prove the converse for $\mathcal{G}_{15}\cup\mathcal{G}_{21}$. In this channel, $R_3$ is upper bounded as
\begin{align}\label{proof51}
&nR_3=H(M_3)=H(M_3 \mid Y_3^{(n)},M_2)+I(M_3;Y_3^{(n)},M_2) \nonumber\\
&\overset{(a)}{=}H(M_3 \mid Y_3^{(n)},M_2)+I(M_3;Y_3^{(n)}\mid M_2) \nonumber\\
&=H(M_3 \mid Y_3^{(n)},M_2)+h(Y_3^{(n)}\mid M_2)-h(Y_3^{(n)}\mid M_2,M_3) \nonumber\\
&\overset{(b)}{\leq} n\epsilon_n+h(Y_3^{(n)}\mid M_2)-h(Y_3^{(n)}\mid M_2,M_3) \nonumber\\
&\overset{(c)}{\leq} n\epsilon_n+\frac{n}{2}\log2\pi e(P+N_3)-h(Y_3^{(n)}\mid M_2,M_3) \nonumber\\
&\overset{(d)}{=}n\epsilon_n+\frac{n}{2}\log2\pi e(P+N_3)-\frac{n}{2}\log2\pi e(\alpha P+N_3),
\end{align}
where (a) follows from the independence of $M_2$ and $M_3$, (b) from \eqref{fano}, (c) from $h(Y_3^{(n)}\mid M_2)\leq h(Y_3^{(n)})\leq \frac{n}{2}\log2\pi e(P+N_3)$, and (d) from the fact that
\begin{multline*}
\frac{n}{2}\log2\pi e N_3=h(Z_3^{(n)})= h(Y_3^{(n)}\mid X^{(n)})\\
\overset{(e)}{\leq} h(Y_3^{(n)}\mid M_2,M_3)\leq h(Y_3^{(n)}) \leq \frac{n}{2}\log2\pi e(P+N_3),
\end{multline*}
where (e) is because $(M_2,M_3)\rightarrow X^{(n)}\rightarrow Y_3^{(n)}$ form a Markov chain; then since $\frac{n}{2}\log2\pi e N_3\leq h(Y_3^{(n)}\mid M_2,M_3)\leq \frac{n}{2}\log2\pi e(P+N_3)$ 
there must exist an $0\leq \alpha \leq1$ such that $h(Y_3^{(n)}\mid M_2,M_3)=\frac{n}{2}\log2\pi e(\alpha P+N_3)$.

$R_2+R_3$ is upper bounded as
\begin{align}\label{proof52}
&n(R_2+R_3)=H(M_2,M_3)\nonumber\\
&=H(M_2, M_3 \mid Y_2^{(n)},M_1)+I(M_2, M_3;Y_2^{(n)},M_1) \nonumber\\
&\overset{(a)}{=}H(M_2,M_3\mid Y_2^{(n)},M_1)+I(M_2, M_3;Y_2^{(n)} \mid M_1) \nonumber\\
&=H(M_2, M_3 \mid Y_2^{(n)},M_1)\nonumber\\
&\hskip40pt+h(Y_2^{(n)}\mid M_1)-h(Y_2^{(n)}\mid M_1,M_2,M_3) \nonumber\\
&\overset{(b)}{\leq}2 n\epsilon_n+h(Y_2^{(n)}\mid M_1)-h(Y_2^{(n)}\mid M_1,M_2,M_3) \nonumber\\
&\overset{(c)}{\leq} 2 n\epsilon_n+\frac{n}{2}\log2\pi e(P+N_2)-h(Y_2^{(n)}\mid M_1,M_2,M_3)\nonumber\\
&\overset{(d)}{\leq} 2 n\epsilon_n+\frac{n}{2}\log2\pi e(P+N_2)-\frac{n}{2}\log2\pi e N_2,
\end{align}
where (a) follows from the independence of $M_1$, $M_2$ and $M_3$, and (b) from adding the following inequalities which are the results of using Lemma \ref{modifiedfano} and \eqref{fano} as
\begin{align*}
&H(M_3 \mid Y_2^{(n)},M_2,M_1)\leq \\
&\hskip50pt H(M_3 \mid Y_2^{(n)},M_2)\leq H(M_3 \mid Y_3^{(n)},M_2) \leq n \epsilon_n,\\
&H(M_2 \mid Y_2^{(n)},M_1) \leq n \epsilon_n.
\end{align*}
In \eqref{proof52}, (c) follows from $h(Y_2^{(n)}\mid M_1)\leq h(Y_2^{(n)})\leq \frac{n}{2}\log2\pi e(P+N_2)$ and (d) from  (\ref{noise2entropy}). 

$R_1$ is upper bounded as
\begin{align}\label{proof53}
&n(R_1)=H(M_1)\nonumber\\
&=H(M_1 \mid Y_1^{(n)},M_2,M_3)+I(M_1;Y_1^{(n)},M_2,M_3) \nonumber\\
&\overset{(a)}{=}H(M_1 \mid Y_1^{(n)},M_2,M_3)+I(M_1;Y_1^{(n)} \mid M_2,M_3) \nonumber\\
&=H(M_1 \mid Y_1^{(n)},M_2,M_3)\nonumber\\
&\hskip40pt+h(Y_1^{(n)}\mid M_2,M_3)-h(Y_1^{(n)}\mid M_1,M_2,M_3) \nonumber\\
&\overset{(b)}{\leq} n\epsilon_n+h(Y_1^{(n)}\mid M_3,M_2)-h(Y_1^{(n)}\mid M_1,M_2,M_3) \nonumber\\
&\overset{(c)}{\leq} n\epsilon_n+\frac{n}{2}\log2\pi e(\alpha P+N_1)-h(Y_1^{(n)}\mid M_1,M_2,M_3)\nonumber\\
&\overset{(d)}{\leq} n\epsilon_n+\frac{n}{2}\log2\pi e(\alpha P+N_1)-\frac{n}{2}\log2\pi e N_1,
\end{align}
where (a) follows from the independence of $M_1$, $M_2$ and $M_3$, (b) from \eqref{fano} and $H(M_1 \mid Y_1^{(n)},M_3,M_2)\leq H(M_1 \mid Y_1^{(n)})$, (c) from Corollary \ref{epicoro} and $h(Y_3^{(n)}\mid M_2,M_3)=\frac{n}{2}\log2\pi e(\alpha P+N_3)$, and (d) from (\ref{noise1entropy}).

$R_1+R_2+R_3$ is also upper bounded as
\begin{align}\label{proof54}
&n(R_1+R_2+R_3)=H(M_1,M_2,M_3)\nonumber\\
&=H(M_1, M_2,M_3 \mid Y_1^{(n)})+I(M_1, M_2,M_3;Y_1^{(n)}) \nonumber\\
&=H(M_1, M_2,M_3 \mid Y_1^{(n)})\nonumber\\
&\hskip40pt+h(Y_1^{(n)})-h(Y_1^{(n)}\mid M_1,M_2,M_3) \nonumber\\
&\overset{(a)}{\leq}3 n\epsilon_n+h(Y_1^{(n)})-h(Y_1^{(n)}\mid M_1,M_2,M_3) \nonumber\\
&\overset{(b)}{\leq} 3 n\epsilon_n+\frac{n}{2}\log2\pi e(P+N_1)-h(Y_1^{(n)}\mid M_1,M_2,M_3)\nonumber\\
&\overset{(c)}{\leq} 3 n\epsilon_n+\frac{n}{2}\log2\pi e(P+N_1)-\frac{n}{2}\log2\pi e N_1,
\end{align}
where (a) follows from adding the following inequalities which are the results of using Lemma \ref{modifiedfano} and \eqref{fano} as
\begin{align*}
&H(M_3 \mid Y_1^{(n)},M_2,M_1)\leq \nonumber\\
&\hskip30pt H(M_3 \mid Y_1^{(n)},M_2)\leq H(M_3 \mid Y_3^{(n)},M_2) \leq n \epsilon_n,\\
&H(M_2 \mid Y_1^{(n)},M_1)\leq H(M_2 \mid Y_2^{(n)},M_1) \leq n \epsilon_n,\\
&H(M_1 \mid Y_1^{(n)}) \leq n \epsilon_n.
\end{align*}
In \eqref{proof54}, (b) follows from $h(Y_1^{(n)})\leq\frac{n}{2}\log2\pi e(P+N_1)$, and (c) from (\ref{noise1entropy}).

From (\ref{proof51})-(\ref{proof54}) and since $\epsilon_n$ goes to zero as $n \rightarrow \infty$, the proof for this member is complete.

The converse proof for the other members of this group except $\mathcal{G}_{15}\cup\mathcal{G}_{22}$ is straightforward; we need to modify \eqref{proof52}, if receiver 2 knows $M_3$, and \eqref{proof54} if receiver 1 knows $M_2$ or $M_3$. 

For $\mathcal{G}_{15}\cup\mathcal{G}_{22}$ (the member with different transmission scheme) in group 5, the given rate region in Table \ref{TransmissionSchemes} can be rewritten as
\begin{align*}
&R_1+R_2 < C\left(\frac{ P}{N_1}\right),\\
&R_1+R_3 < C\left(\frac{ P}{N_1}\right),\\
&R_1 < C\left(\frac{\alpha P}{N_1}\right),\\
&R_2 < C\left(\frac{ P}{N_2}\right),\\
&R_3 < C\left(\frac{(1-\alpha)P}{\alpha P+N_3}\right),
\end{align*}
where $0\leq\alpha\leq1$. The achievability of this region using the transmission scheme, given in Table \ref{TransmissionSchemes} for $\mathcal{G}_{15}\cup\mathcal{G}_{22}$, can be verified by using successive decoding at receiver 3 and simultaneous decoding at receivers 1 and 2. We also need to know that $m_2\oplus m_3$  from the standpoint of receiver 3 has the unknown information rate of $R_3$, from the standpoint of receiver 2 has the unknown information rate of  $R_2$ and from the standpoint of receiver 1 has the unknown information rate of $\max\{R_2,R_3\}$.

Here, we prove the converse for $\mathcal{G}_{15}\cup\mathcal{G}_{22}$. In this channel, $R_3$ is upper bounded as
\begin{align}\label{proof55}
&nR_3=H(M_3)=H(M_3 \mid Y_3^{(n)},M_2)+I(M_3;Y_3^{(n)},M_2) \nonumber\\
&\overset{(a)}{=}H(M_3 \mid Y_3^{(n)},M_2)+I(M_3;Y_3^{(n)}\mid M_2) \nonumber\\
&=H(M_3 \mid Y_3^{(n)},M_2)+h(Y_3^{(n)}\mid M_2)-h(Y_3^{(n)}\mid M_2,M_3) \nonumber\\
&\overset{(b)}{\leq} n\epsilon_n+h(Y_3^{(n)}\mid M_2)-h(Y_3^{(n)}\mid M_2,M_3) \nonumber\\
&\overset{(c)}{\leq} n\epsilon_n+\frac{n}{2}\log2\pi e(P+N_3)-h(Y_3^{(n)}\mid M_2,M_3) \nonumber\\
&\overset{(d)}{=}n\epsilon_n+\frac{n}{2}\log2\pi e(P+N_3)-\frac{n}{2}\log2\pi e(\alpha P+N_3),
\end{align}
where (a) follows from the independence of $M_2$ and $M_3$, (b) from \eqref{fano}, (c) from $h(Y_3^{(n)}\mid M_2)\leq h(Y_3^{(n)})\leq \frac{n}{2}\log2\pi e(P+N_3)$, and (d) from the fact that
\begin{multline*}
\frac{n}{2}\log2\pi e N_3=h(Z_3^{(n)})= h(Y_3^{(n)}\mid X^{(n)})\\
\overset{(e)}{\leq} h(Y_3^{(n)}\mid M_2,M_3)\leq h(Y_3^{(n)}) \leq \frac{n}{2}\log2\pi e(P+N_3),
\end{multline*}
where (e) is because $(M_2,M_3)\rightarrow X^{(n)}\rightarrow Y_3^{(n)}$ form a Markov chain; then since $\frac{n}{2}\log2\pi e N_3\leq h(Y_3^{(n)}\mid M_2,M_3)\leq \frac{n}{2}\log2\pi e(P+N_3)$ there must exist an $0\leq \alpha \leq1$ such that $h(Y_3^{(n)}\mid M_3,M_2)=\frac{n}{2}\log2\pi e(\alpha P+N_3)$. 

$R_2$ is upper bounded as
\begin{align}\label{proof56}
&n R_2=H(M_2)\nonumber\\
&=H(M_2 \mid Y_2^{(n)},M_1,M_3)+I(M_2;Y_2^{(n)},M_1,M_3) \nonumber\\
&\overset{(a)}{=}H(M_2\mid Y_2^{(n)},M_1,M_3)+I(M_2;Y_2^{(n)} \mid M_1, M_3) \nonumber\\
&=H(M_2 \mid Y_2^{(n)},M_1,M_3)\nonumber\\
&\hskip40pt+h(Y_2^{(n)}\mid M_1,M_3)-h(Y_2^{(n)}\mid M_1,M_2,M_3) \nonumber\\
&\overset{(b)}{\leq}n\epsilon_n+h(Y_2^{(n)}\mid M_1, M_3)-h(Y_2^{(n)}\mid M_1,M_2,M_3) \nonumber\\
&\overset{(c)}{\leq} n\epsilon_n+\frac{n}{2}\log2\pi e(P+N_2)-h(Y_2^{(n)}\mid M_1,M_2,M_3)\nonumber\\
&\overset{(d)}{\leq} n\epsilon_n+\frac{n}{2}\log2\pi e(P+N_2)-\frac{n}{2}\log2\pi e N_2,
\end{align}
where (a) follows from the independence of $M_1$, $M_2$ and $M_3$, (b) from \eqref{fano}, (c) from $h(Y_2^{(n)}\mid M_1,M_3)\leq h(Y_2^{(n)})\leq\frac{n}{2}\log2\pi e(P+N_2)$, and (d) from (\ref{noise2entropy}). 

$R_1$ is upper bounded as
\begin{align}\label{proof57}
&nR_1=H(M_1)\nonumber\\
&=H(M_1 \mid Y_1^{(n)},M_2,M_3)+I(M_1;Y_1^{(n)},M_2,M_3) \nonumber\\
&\overset{(a)}{=}H(M_1 \mid Y_1^{(n)},M_2,M_3)+I(M_1;Y_1^{(n)} \mid M_2,M_3) \nonumber\\
&=H(M_1 \mid Y_1^{(n)},M_2,M_3)\nonumber\\
&\hskip40pt+h(Y_1^{(n)}\mid M_2,M_3)-h(Y_1^{(n)}\mid M_1,M_2,M_3) \nonumber\\
&\overset{(b)}{\leq} n\epsilon_n+h(Y_1^{(n)}\mid M_2,M_3)-h(Y_1^{(n)}\mid M_1,M_2,M_3) \nonumber\\
&\overset{(c)}{\leq} n\epsilon_n+\frac{n}{2}\log2\pi e(\alpha P+N_1)-h(Y_1^{(n)}\mid M_1,M_2,M_3)\nonumber\\
&\overset{(d)}{\leq} n\epsilon_n+\frac{n}{2}\log2\pi e(\alpha P+N_1)-\frac{n}{2}\log2\pi e N_1,
\end{align}
where (a) follows from the independence of $M_1$, $M_2$ and $M_3$, (b) from \eqref{fano} and $H(M_1 \mid Y_1^{(n)},M_3,M_2)\leq H(M_1 \mid Y_1^{(n)})$, (c) from Corollary \ref{epicoro} and $h(Y_3^{(n)}\mid M_3,M_2)=\frac{n}{2}\log2\pi e(\alpha P+N_3)$, and (d) from (\ref{noise1entropy}).

$R_1+R_2$ is upper bounded as
\begin{align}\label{proof58}
&n(R_1+R_2)=H(M_1,M_2)\nonumber\\
&=H(M_1,M_2 \mid Y_1^{(n)},M_3)+I(M_1,M_2;Y_1^{(n)},M_3) \nonumber\\
&\overset{(a)}{=}H(M_1,M_2 \mid Y_1^{(n)},M_3)+I(M_1,M_2;Y_1^{(n)}\mid M_3) \nonumber\\
&=H(M_1,M_2 \mid Y_1^{(n)},M_3)\nonumber\\
&\hskip40pt+h(Y_1^{(n)}\mid M_3)-h(Y_1^{(n)}\mid M_1,M_2,M_3) \nonumber\\
&\overset{(b)}{\leq}2 n\epsilon_n+h(Y_1^{(n)}\mid M_3)-h(Y_1^{(n)}\mid M_1,M_2,M_3) \nonumber\\
&\overset{(c)}{\leq} 2 n\epsilon_n+\frac{n}{2}\log2\pi e(P+N_1)-h(Y_1^{(n)}\mid M_1,M_2,M_3)\nonumber\\
&\overset{(d)}{\leq} 2 n\epsilon_n+\frac{n}{2}\log2\pi e(P+N_1)-\frac{n}{2}\log2\pi e N_1,
\end{align}
where (a) follows from the independence of $M_1$, $M_2$ and $M_3$, and (b) from adding the following inequalities which are the results of using Lemma \ref{modifiedfano} and \eqref{fano} as
\begin{align*}
&H(M_2 \mid Y_1^{(n)},M_3,M_1)\leq H(M_2 \mid Y_2^{(n)},M_3,M_1) \leq n \epsilon_n,\\
&H(M_1 \mid Y_1^{(n)},M_3) \leq H(M_1 \mid Y_1^{(n)})\leq n \epsilon_n.
\end{align*}
In \eqref{proof58}, (c) follows from $h(Y_1^{(n)}\mid M_3)\leq h(Y_1^{(n)})\leq \frac{n}{2}\log2\pi e(P+N_1)$ and (d) from (\ref{noise1entropy}).

$R_1+R_3$ is also upper bounded as
\begin{align}\label{proof59}
&n(R_1+R_3)=H(M_1,M_3)\nonumber\\
&=H(M_1,M_3 \mid Y_1^{(n)},M_2)+I(M_1,M_3;Y_1^{(n)},M_2) \nonumber\\
&\overset{(a)}{=}H(M_1,M_3 \mid Y_1^{(n)},M_2)+I(M_1,M_3;Y_1^{(n)}\mid M_2) \nonumber\\
&=H(M_1,M_3 \mid Y_1^{(n)},M_2)\nonumber\\
&\hskip40pt+h(Y_1^{(n)}\mid M_2)-h(Y_1^{(n)}\mid M_1,M_2,M_3) \nonumber\\
&\overset{(b)}{\leq}2 n\epsilon_n+h(Y_1^{(n)}\mid M_2)-h(Y_1^{(n)}\mid M_1,M_2,M_3) \nonumber\\
&\overset{(c)}{\leq} 2 n\epsilon_n+\frac{n}{2}\log2\pi e(P+N_1)-h(Y_1^{(n)}\mid M_1,M_2,M_3)\nonumber\\
&\overset{(d)}{\leq} 2 n\epsilon_n+\frac{n}{2}\log2\pi e(P+N_1)-\frac{n}{2}\log2\pi e N_1,
\end{align}
where (a) follows from the independence of $M_1$, $M_2$ and $M_3$, and (b) from adding the following inequalities which are the results of using Lemma \ref{modifiedfano} and \eqref{fano} as
\begin{align*}
&H(M_3 \mid Y_1^{(n)},M_2,M_1)\leq H(M_3 \mid Y_1^{(n)},M_2)\\
&\hskip100pt\leq H(M_3 \mid Y_3^{(n)},M_2) \leq n \epsilon_n,\\
&H(M_1 \mid Y_1^{(n)},M_2) \leq H(M_1 \mid Y_1^{(n)})\leq n \epsilon_n.
\end{align*}
In \eqref{proof59}, (c) follows from $h(Y_1^{(n)}\mid M_2)\leq h(Y_1^{(n)})\leq \frac{n}{2}\log2\pi e(P+N_1)$ and (d) from (\ref{noise1entropy}).

From (\ref{proof55})-(\ref{proof59}) and since $\epsilon_n$ goes to zero as $n \rightarrow \infty$, the converse proof for $\mathcal{G}_{15}\cup\mathcal{G}_{22}$ is complete.

Group 6: The achievability for $\mathcal{G}_{16} \cup \mathcal{G}_{21}$ is proved using successive decoding at receivers 2 and 3 and simultaneous decoding at receiver 1. Since receivers 2 and 3 know $M_1$, the second and the third inequalities, given in Table \ref{TransmissionSchemes} for this group, are required for achievability. Receiver 1, using simultaneous decoding, decodes $\hat{m}_1$ if there exists a unique $\hat{m}_1$ such that $(X_1^{(n)}([\hat{m}_1,m_2]),X_2^{(n)}([\hat{m}_1,m_3]),Y_1^{(n)})\in \mathcal{T}_\delta^{(n)}$ for some $m_2$, $m_3$; otherwise the error is declared. Assuming the transmitted messages are equal to one by the symmetry of the code generation, the error events at receiver 1 for $\mathcal{G}_{16} \cup \mathcal{G}_{21}$ are
\begin{align*}
&\mathcal{E}_{11}:\left(X_1^{(n)}([1,m_2]),X_2^{(n)}([1,m_3]),Y_1^{(n)}\right)\notin \mathcal{T}_\delta^{(n)}\\ 
&\hskip175pt\mathrm{for\;all}\;m_2, m_3,\\
&\mathcal{E}_{12}:\left(X_1^{(n)}([m_1,m_2]),X_2^{(n)}([m_1,m_3]),Y_1^{(n)}\right)\in \mathcal{T}_\delta^{(n)}\\ 
&\hskip130pt \text{for some}\;m_1\neq1, m_2,  m_3.
\end{align*}
According to these error events, it can be seen for $\mathcal{G}_{16}\cup\mathcal{G}_{21}$, the first inequality for this group, $R_1+R_2+R_3<C(P/N_1)$, guarantees that the probability of error at receiver 1 tends to zero as $n$ increases.

For all other elements in group 6, we use the same encoding and decoding schemes, but each receiver makes its decoding decision based on its extra side information.

Here, we prove the converse for $\mathcal{G}_{16} \cup \mathcal{G}_{21}$. The rate $R_3$ in this channel is upper bounded as
\vspace{-0pt}
\begin{align}\label{proof61}
&nR_3=H(M_3)=H(M_3 \mid Y_3^{(n)},M_1)+I(M_3;Y_3^{(n)},M_1) \nonumber\\
&\overset{(a)}{=}H(M_3 \mid Y_3^{(n)},M_1)+I(M_3;Y_3^{(n)}\mid M_1) \nonumber\\
&=H(M_3 \mid Y_3^{(n)},M_1)+h(Y_3^{(n)}\mid M_1)-h(Y_3^{(n)}\mid M_1, M_3) \nonumber\\
&\overset{(b)}{\leq} n\epsilon_n+h(Y_3^{(n)}\mid M_1)-h(Y_3^{(n)}\mid M_1,M_3) \nonumber\\
&\overset{(c)}{\leq} n\epsilon_n+\frac{n}{2}\log2\pi e(P+N_3)-h(Y_3^{(n)}\mid M_1,M_3) \nonumber\\
&\overset{(d)}{=}n\epsilon_n+\frac{n}{2}\log2\pi e(P+N_3)-\frac{n}{2}\log2\pi e (\alpha P+N_3),
\end{align}
where (a) follows from the independence of $M_1$ and $M_3$, (b) from \eqref{fano}, (c) from $h(Y_3^{(n)}\mid M_1)\leq h(Y_3^{(n)})\leq \frac{n}{2}\log2\pi e(P+N_3)$, and (d) from the fact that
\vspace{-0pt}
\begin{align*}
&\frac{n}{2}\log2\pi e N_3=h(Z_3^{(n)})= h(Y_3^{(n)}\mid X^{(n)})\\
&\overset{(e)}{\leq} h(Y_3^{(n)}\mid M_1,M_3)\leq h(Y_3^{(n)}) \leq \frac{n}{2}\log2\pi e(P+N_3),
\end{align*}
where (e) is because $(M_1,M_3)\rightarrow X^{(n)} \rightarrow Y_3^{(n)}$ form a Markov chain; then since $\frac{n}{2}\log2\pi e N_3\leq h(Y_3^{(n)}\mid M_1,M_ 3)\leq \frac{n}{2}\log2\pi e(P+N_3)$ there must exist an $0\leq \alpha \leq1$ such that $h(Y_3^{(n)}\mid M_1,M_3)=\frac{n}{2}\log2\pi e(\alpha P+N_3)$.

$R_2$ is restricted from the above as
\vspace{-0pt}
\begin{align}\label{proof62}
&nR_2=H(M_2)\nonumber \\
&=H(M_2 \mid Y_2^{(n)},M_1,M_3)+I(M_2;Y_2^{(n)},M_1,M_3) \nonumber\\
&\overset{(a)}{=}H(M_2 \mid Y_2^{(n)},M_1,M_3)+I(M_2;Y_2^{(n)} \mid M_1,M_3) \nonumber\\
&=H(M_2 \mid Y_2^{(n)},M_1,M_3)\nonumber\\
&\hskip40pt+h(Y_2^{(n)}\mid M_1,M_3)-h(Y_2^{(n)}\mid M_1,M_2,M_3) \nonumber\\
&\overset{(b)}{\leq}n\epsilon_n+h(Y_2^{(n)}\mid M_1,M_3)-h(Y_2^{(n)}\mid M_1,M_2,M_3) \nonumber\\
&\overset{(c)}{\leq} n\epsilon_n+\frac{n}{2}\log2\pi e(\alpha P+N_2)-h(Y_2^{(n)}\mid M_1,M_2,M_3)\nonumber\\
&\overset{(d)}{\leq} n\epsilon_n+\frac{n}{2}\log2\pi e(\alpha P+N_2)-\frac{n}{2}\log2\pi e N_2,
\end{align}
where (a) follows from the independence of $M_1$, $M_2$ and $M_3$, (b) from \eqref{fano} and $H(M_2 \mid Y_2^{(n)},M_1,M_3)\leq H(M_2 \mid Y_2^{(n)},M_1)$, (c) from Corollary \ref{epicoro} and $h(Y_3^{(n)}\mid M_1,M_3)=\frac{n}{2}\log2\pi e(\alpha P+N_3)$, and (d) from (\ref{noise2entropy}). 

For this channel, we also have
\begin{align}\label{proof63}
&n(R_1+R_2+R_3)=H(M_1, M_2,M_3)\nonumber\\
&=H(M_1, M_2, M_3 \mid Y_1^{(n)})+I(M_1, M_2, M_3;Y_1^{(n)}) \nonumber\\
&=H(M_1,M_2, M_3 \mid Y_1^{(n)})\nonumber\\
&\hskip80pt+h(Y_1^{(n)})-h(Y_1^{(n)}\mid M_1,M_2,M_3) \nonumber \\
&\overset{(a)}{\leq}3 n\epsilon_n+h(Y_1^{(n)})-h(Y_1^{(n)}\mid M_1,M_2,M_3) \nonumber\\
&\overset{(b)}{\leq} 3 n\epsilon_n+\frac{n}{2}\log2\pi e(P+N_1)-h(Y_1^{(n)}\mid M_1,M_2,M_3)\nonumber\\
&\overset{(c)}{\leq} 3 n\epsilon_n+\frac{n}{2}\log2\pi e(P+N_1)-\frac{n}{2}\log2\pi e N_1,
\end{align}
where (a) follows from adding the following inequalities which are the results of using Lemma \ref{modifiedfano} and \eqref{fano} as
\vspace{-2pt}
\begin{align*}
&H(M_1 \mid Y_1^{(n)}) \leq n \epsilon_n,\\
&H(M_2 \mid Y_1^{(n)},M_1) \leq H(M_2 \mid Y_2^{(n)},M_1) \leq n \epsilon_n,\\
&H(M_3 \mid Y_1^{(n)},M_1,M_2)\leq H(M_3 \mid Y_1^{(n)},M_1)\\
&\hskip110pt\leq H(M_3 \mid Y_3^{(n)},M_1) \leq n \epsilon_n.
\end{align*}
In (\ref{proof63}), (b) follows from $h(Y_1^{(n)})\leq\frac{n}{2}\log2\pi e(P+N_1)$, and (c) from (\ref{noise1entropy}).

From (\ref{proof61})--(\ref{proof63}) and since $\epsilon_n \rightarrow 0$ as $n \rightarrow \infty$, the converse for $\mathcal{G}_{16} \cup \mathcal{G}_{21}$ is proven. The converse for the other channels  is straightforward; we only need to modify \eqref{proof63}, if receiver 1 knows $M_2$ or $M_3$.

Group 8: The achievability of the given rate region for this group is proved by considering the fact that each receiver decodes the correct $x^{(n)}$ over the set of valid candidates which is determined based on its side information. 

Here, we prove the converse for $\mathcal{G}_{18}\cup\mathcal{G}_{21}$. In this channel $R_3$ is upper bounded as
\begin{align}\label{proof81}
&nR_3=H(M_3)\nonumber\\
&=H(M_3 \mid Y_3^{(n)},M_1, M_2)+I(M_3;Y_3^{(n)},M_2, M_1) \nonumber\\
&\overset{(a)}{=}H(M_3 \mid Y_3^{(n)},M_1, M_2)+I(M_3;Y_3^{(n)}\mid M_1, M_2) \nonumber\\
&=H(M_3 \mid Y_3^{(n)},M_1, M_2)\nonumber\\
&\hskip40pt+h(Y_3^{(n)}\mid M_1, M_2)-h(Y_3^{(n)}\mid M_1, M_2, M_3) \nonumber\\
&\overset{(b)}{\leq} n\epsilon_n+h(Y_3^{(n)}\mid M_1,M_2)-h(Y_3^{(n)}\mid M_1,M_2,M_3) \nonumber\\
&\overset{(c)}{\leq} n\epsilon_n+\frac{n}{2}\log2\pi e(P+N_3)-h(Y_3^{(n)}\mid M_1,M_2,M_3) \nonumber\\
&\overset{(d)}{\leq}n\epsilon_n+\frac{n}{2}\log2\pi e(P+N_3)-\frac{n}{2}\log2\pi e N_3,
\end{align}
where (a) follows from the independence of $M_1$, $M_2$ and $M_3$, (b) from \eqref{fano}, (c) from $h(Y_3^{(n)}\mid M_2,M_1)\leq h(Y_3^{(n)})\leq \frac{n}{2}\log2\pi e(P+N_3)$ and (d) from
\begin{multline}\label{noise3entropy}
h(Y_3^{(n)}\mid M_1,M_2,M_3)\\
\overset{(e)}{\geq}(Y_3^{(n)}\mid X^{(n)})=h(Z_3^{(n)})=\frac{n}{2}\log2\pi e N_3,
\end{multline}
where (e) is because $(M_1,M_2,M_3)\rightarrow X^{(n)}\rightarrow Y_3^{(n)}$ form a Markov chain.
For this channel, we also have
\begin{align}\label{proof82}
&n(R_2+R_3)=H(M_2,M_3)\nonumber\\
&=H(M_2, M_3 \mid Y_2^{(n)},M_1)+I(M_2, M_3;Y_2^{(n)},M_1) \nonumber\\
&\overset{(a)}{=}H(M_2,M_3\mid Y_2^{(n)},M_1)+I(M_2, M_3;Y_2^{(n)} \mid M_1) \nonumber\\
&=H(M_2, M_3 \mid Y_2^{(n)},M_1)\nonumber\\
&\hskip40pt+h(Y_2^{(n)}\mid M_1)-h(Y_2^{(n)}\mid M_1,M_2,M_3) \nonumber\\
&\overset{(b)}{\leq}2 n\epsilon_n+h(Y_2^{(n)}\mid M_1)-h(Y_2^{(n)}\mid M_1,M_2,M_3) \nonumber\\
&\overset{(c)}{\leq} 2 n\epsilon_n+\frac{n}{2}\log2\pi e(P+N_2)-h(Y_2^{(n)}\mid M_1,M_2,M_3)\nonumber\\
&\overset{(d)}{\leq} 2 n\epsilon_n+\frac{n}{2}\log2\pi e(P+N_2)-\frac{n}{2}\log2\pi e N_2,
\end{align}
where (a) follows from the independence of $M_1$, $M_2$ and $M_3$, and (b) from adding the following inequalities which are the results of using Lemma \ref{modifiedfano} and \eqref{fano} as
\begin{align*}
&H(M_3 \mid Y_2^{(n)},M_1,M_2)\leq H(M_3 \mid Y_3^{(n)},M_1,M_2) \leq n \epsilon_n,\\
&H(M_2 \mid Y_2^{(n)},M_1) \leq n \epsilon_n.
\end{align*}
In \eqref{proof82}, (c) follows from $h(Y_2^{(n)}\mid M_1)\leq h(Y_2^{(n)})\leq \frac{n}{2}\log2\pi e(P+N_2)$ and (d) from (\ref{noise2entropy}). Finally, for this channel we have
\begin{align}\label{proof83}
&n(R_1+R_2+R_3)=H(M_1, M_2,M_3)\nonumber\\
&=H(M_1, M_2, M_3 \mid Y_1^{(n)})+I(M_1, M_2, M_3;Y_1^{(n)}) \nonumber\\
&=H(M_1,M_2, M_3 \mid Y_1^{(n)})\nonumber\\
&\hskip80pt+h(Y_1^{(n)})-h(Y_1^{(n)}\mid M_1,M_2,M_3) \nonumber\\
&\overset{(a)}{\leq}3 n\epsilon_n+h(Y_1^{(n)})-h(Y_1^{(n)}\mid M_1,M_2,M_3) \nonumber\\
&\overset{(b)}{\leq} 3 n\epsilon_n+\frac{n}{2}\log2\pi e(P+N_1)-h(Y_1^{(n)}\mid M_1,M_2,M_3)\nonumber\\
&\overset{(c)}{\leq} 3 n\epsilon_n+\frac{n}{2}\log2\pi e(P+N_1)-\frac{n}{2}\log2\pi e N_1,
\end{align}
where (a) follows from adding the following inequalities which are the results of using Lemma \ref{modifiedfano} and \eqref{fano} as
\begin{align*}
&H(M_3 \mid Y_1^{(n)},M_1,M_2)\leq H(M_3 \mid Y_3^{(n)},M_1,M_2) \leq n \epsilon_n,\\
&H(M_2 \mid Y_1^{(n)},M_1)\leq H(M_2 \mid Y_2^{(n)},M_1) \leq n \epsilon_n,\\
&H(M_1 \mid Y_1^{(n)}) \leq n \epsilon_n.
\end{align*}
In \eqref{proof83}, (b) follows from $h(Y_1^{(n)})\leq\frac{n}{2}\log2\pi e(P+N_1)$, and (c) from (\ref{noise1entropy}).

From (\ref{proof81}), (\ref{proof82}), (\ref{proof83}) and since $\epsilon_n$ goes to zero as $n \rightarrow \infty$, the converse proof for this member is complete. The converse proof for the other members except $\mathcal{G}_{18}\cup\mathcal{G}_{22}$ is straightforward; we need to modify \eqref{proof82} if receiver 2 knows $M_3$, and \eqref{proof83} if receiver 1 knows $M_2$ or $M_3$. 

Here, we prove the converse for $\mathcal{G}_{18}\cup\mathcal{G}_{22}$ (the member with different transmission scheme). For this member, the given rate region in Table \ref{TransmissionSchemes} can be rewritten as
\begin{align*}
&R_1+R_2\leq C\left(\frac{P}{N_1}\right),\\
&R_1+R_3\leq C\left(\frac{P}{N_1}\right),\\
&R_2\leq C\left(\frac{P}{N_2}\right),\\
&R_3\leq C\left(\frac{P}{N_3}\right).
\end{align*}
In this channel, $R_3$ is upper bounded as
\begin{align}\label{proof84}
&nR_3=H(M_3)\nonumber\\
&=H(M_3 \mid Y_3^{(n)},M_1,M_2)+I(M_3;Y_3^{(n)},M_1,M_2) \nonumber\\
&\overset{(a)}{=}H(M_3 \mid Y_3^{(n)},M_1,M_2)+I(M_3;Y_3^{(n)}\mid M_1,M_2) \nonumber\\
&=H(M_3 \mid Y_3^{(n)},M_1,M_2)\nonumber\\
&\hskip40pt+h(Y_3^{(n)}\mid M_1,M_2)-h(Y_3^{(n)}\mid M_1,M_2,M_3) \nonumber\\
&\overset{(b)}{\leq} n\epsilon_n+h(Y_3^{(n)}\mid M_1,M_2)-h(Y_3^{(n)}\mid M_1,M_2,M_3) \nonumber\\
&\overset{(c)}{\leq} n\epsilon_n+\frac{n}{2}\log2\pi e(P+N_3)-h(Y_3^{(n)}\mid M_1,M_2,M_3) \nonumber\\
&\overset{(d)}{\leq}n\epsilon_n+\frac{n}{2}\log2\pi e(P+N_3)-\frac{n}{2}\log2\pi e N_3,
\end{align}
where (a) follows from the independence of $M_1$, $M_2$ and $M_3$, (b) from \eqref{fano}, (c) from $h(Y_3^{(n)}\mid M_1,M_2)\leq h(Y_3^{(n)})\leq \frac{n}{2}\log2\pi e(P+N_3)$ and (d) from  (\ref{noise3entropy}). 

$R_2$ is upper bounded as 
\begin{align}\label{proof85}
&n R_2=H(M_2)\nonumber\\
&=H(M_2 \mid Y_2^{(n)},M_1,M_3)+I(M_2;Y_2^{(n)},M_1,M_3) \nonumber\\
&\overset{(a)}{=}H(M_2\mid Y_2^{(n)},M_1,M_3)+I(M_2;Y_2^{(n)} \mid M_1, M_3) \nonumber\\
&=H(M_2 \mid Y_2^{(n)},M_1,M_3)\nonumber\\
&\hskip40pt+h(Y_2^{(n)}\mid M_1,M_3)-h(Y_2^{(n)}\mid M_1,M_2,M_3) \nonumber\\
&\overset{(b)}{\leq}n\epsilon_n+h(Y_2^{(n)}\mid M_1, M_3)-h(Y_2^{(n)}\mid M_1,M_2,M_3) \nonumber\\
&\overset{(c)}{\leq} n\epsilon_n+\frac{n}{2}\log2\pi e(P+N_2)-h(Y_2^{(n)}\mid M_1,M_2,M_3)\nonumber\\
&\overset{(d)}{\leq} n\epsilon_n+\frac{n}{2}\log2\pi e(P+N_2)-\frac{n}{2}\log2\pi e N_2,
\end{align}
where (a) follows from the independence of $M_1$, $M_2$ and $M_3$, (b) from \eqref{fano}, (c) from $h(Y_2^{(n)}\mid M_1,M_3)\leq h(Y_2^{(n)})\leq \frac{n}{2}\log2\pi e(P+N_2)$ and (d) from (\ref{noise2entropy}). 

$R_1+R_2$ is restricted from the above as
\begin{align}\label{proof86}
&n(R_1+R_2)=H(M_1,M_2)\nonumber\\
&=H(M_1,M_2 \mid Y_1^{(n)},M_3)+I(M_1,M_2;Y_1^{(n)},M_3) \nonumber\\
&\overset{(a)}{=}H(M_1,M_2 \mid Y_1^{(n)},M_3)+I(M_1,M_2;Y_1^{(n)}\mid M_3) \nonumber\\
&=H(M_1,M_2 \mid Y_1^{(n)},M_3)\nonumber\\
&\hskip40pt+h(Y_1^{(n)}\mid M_3)-h(Y_1^{(n)}\mid M_1,M_2,M_3) \nonumber\\
&\overset{(b)}{\leq}2 n\epsilon_n+h(Y_1^{(n)}\mid M_3)-h(Y_1^{(n)}\mid M_1,M_2,M_3) \nonumber\\
&\overset{(c)}{\leq} 2 n\epsilon_n+\frac{n}{2}\log2\pi e(P+N_1)-h(Y_1^{(n)}\mid M_1,M_2,M_3)\nonumber\\
&\overset{(d)}{\leq} 2 n\epsilon_n+\frac{n}{2}\log2\pi e(P+N_1)-\frac{n}{2}\log2\pi e N_1,
\end{align}
where (a) follows from the independence of $M_1$, $M_2$ and $M_3$, (b) from adding the following inequalities which are the results of using Lemma \ref{modifiedfano} and \eqref{fano} as
\begin{align*}
&H(M_2 \mid Y_1^{(n)},M_1,M_3)\leq H(M_2 \mid Y_2^{(n)},M_1,M_3) \leq n \epsilon_n,\\
&H(M_1 \mid Y_1^{(n)},M_3) \leq H(M_1 \mid Y_1^{(n)})\leq n \epsilon_n.
\end{align*}
In \eqref{proof86}, (c) follows from $h(Y_1^{(n)}\mid M_3)\leq h(Y_1^{(n)})\leq \frac{n}{2}\log2\pi e(P+N_1)$ and (d) from (\ref{noise1entropy}). 

Finally for this channel, we have
\begin{align}\label{proof87}
&n(R_1+R_3)=H(M_1,M_3)\nonumber\\
&=H(M_1,M_3 \mid Y_1^{(n)},M_2)+I(M_1,M_3;Y_1^{(n)},M_2) \nonumber\\
&\overset{(a)}{=}H(M_1,M_3 \mid Y_1^{(n)},M_2)+I(M_1,M_3;Y_1^{(n)}\mid M_2) \nonumber\\
&=H(M_1,M_3 \mid Y_1^{(n)},M_2)\nonumber\\
&\hskip40pt+h(Y_1^{(n)}\mid M_2)-h(Y_1^{(n)}\mid M_1,M_2,M_3) \nonumber\\
&\overset{(b)}{\leq}2 n\epsilon_n+h(Y_1^{(n)}\mid M_2)-h(Y_1^{(n)}\mid M_1,M_2,M_3) \nonumber\\
&\overset{(c)}{\leq} 2 n\epsilon_n+\frac{n}{2}\log2\pi e(P+N_1)-h(Y_1^{(n)}\mid M_1,M_2,M_3)\nonumber\\
&\overset{(d)}{\leq} 2 n\epsilon_n+\frac{n}{2}\log2\pi e(P+N_1)-\frac{n}{2}\log2\pi e N_1,
\end{align}
where (a) follows from the independence of $M_1$, $M_2$ and $M_3$, (b) from adding the following inequalities which are the results of using Lemma \ref{modifiedfano} and \eqref{fano} as
\begin{align*}
&H(M_3 \mid Y_1^{(n)},M_2,M_1)\leq H(M_3 \mid Y_3^{(n)},M_2,M_1) \leq n \epsilon_n,\\
&H(M_1 \mid Y_1^{(n)},M_2) \leq H(M_1 \mid Y_1^{(n)})\leq n \epsilon_n.
\end{align*}
In \eqref{proof87}, (c) follows from $h(Y_1^{(n)}\mid M_2)\leq h(Y_1^{(n)})\leq \frac{n}{2}\log2\pi e(P+N_1)$ and (d) from (\ref{noise1entropy}).

From (\ref{proof84})--(\ref{proof87}) and since $\epsilon_n$ goes to zero as $n \rightarrow \infty$, the converse proof for this member is complete.
\end{IEEEproof}

\bibliographystyle{IEEEtran}
%

\end{document}